\documentclass[12pt]{article}

\usepackage[dvips]{color}                 %%% for colors in the graphs
\usepackage{graphicx}                     %%% for pictures and figures
\usepackage{amssymb}
\usepackage{amsmath}

\usepackage{xspace}                       %%% to get the correct spacing of
\renewcommand{\today}{\ifcase\day\or 1st\or 2nd\or 3rd\or 4th\or 5th\or 6th\or
  7th\or 8th\or 9th\or 10th\or 11th\or 12th\or 13th\or 14th\or 15th\or 16th\or
  17th\or 18th\or 19th\or 20th\or 21st\or 22nd\or 23rd\or 24th\or 25th\or
  26th\or 27th\or 28th\or 29th\or 30th\or 31st\fi~\ifcase\month\or January\or
  February\or March\or April\or May\or June\or July\or August\or September\or
  October\or November\or December\fi \space \number\year}

\newcommand{\journal}[4]{{#1}\textbf{#2},
  #4 (#3)}
\newcommand{\EPJA}{\textit{Eur.\  Phys.\ J.\ }\textbf{A}}

\newcommand{\NPA}{\textit{Nucl.\ Phys.\ }\textbf{A}}
\newcommand{\NPB}{\textit{Nucl.\ Phys.\ }\textbf{B}}
\newcommand{\PLB}{\textit{Phys.\ Lett.\ }\textbf{B}}
\newcommand{\PR}{\textit{Phys.\ Rev.\ }}

\newcommand{\PRC}{\PR\textbf{C}}
\newcommand{\PRD}{\PR\textbf{D}}
\newcommand{\PRL}{\PR\textit{Lett.\ }}

\newcommand{\dis}{\displaystyle}
\newcommand{\non}{\nonumber}
\newcommand{\hq}{\hspace{0.5em}}

\newcommand{\half}{\frac{1}{2}}

\newcommand{\kv}{\vec{k}}

\newcommand{\MeV}{\mathrm{MeV}}

\renewcommand{\Re}{\mathrm{Re}}
\renewcommand{\Im}{\mathrm{Im}}

\newcommand{\mytitle}[1]{\begin{center}\LARGE{\textbf{#1}}\end{center}}
\newcommand{\myauthor}[1]{\textbf{#1}}
\newcommand{\myaddress}[1]{\textit{#1}}
\newcommand{\mypreprint}[1]{\begin{flushright}#1\end{flushright}}

\begin{document}
%%%%%%%%%%%%%%%%%%%%%%%%%%%%%%%%%%%%%%%%%%%%%%%%%%%%%%%%%%%%%%%%%%%%%%%%%%%%%%%
% This is a nice title page including abstract ....
%

\begin{titlepage}
  
%  \mypreprint{ \textbf{Draft version \today}\hfill
  \mypreprint{\hfill
    nucl-th/0307070\\
    TUM-T39-03-10\\
    ECT*-03-22\\}

% nucl-th raus, Draft version raus

  %\vspace*{0.5cm}
  \vspace*{0.1cm}
  
  \mytitle{Signatures of Chiral Dynamics in Low Energy Compton Scattering off
    the Nucleon}

  %\vspace*{0.5cm}
  \vspace*{0.3cm}

\begin{center}
  \myauthor{Robert P. Hildebrandt$^{a,}$}\footnote{Email:
    rhildebr@physik.tu-muenchen.de;
    permanent address: a}, \myauthor{Harald W.\ 
    Grie{\ss}hammer$^{a,b,}$}\footnote{Email: hgrie@physik.tu-muenchen.de;
    permanent address: a}, \myauthor{Thomas R.~Hemmert$^{a,b,}$}\footnote{Email:
    themmert@physik.tu-muenchen.de;
    permanent address: a}\\[2ex]
  
  and\\[2ex]
  
  \myauthor{Barbara Pasquini$^{b,c,}$}\footnote{Email: pasquini@pv.infn.it;
    permanent address: c}
  
  \vspace*{0.5cm}
  
  \myaddress{$^a$ Theoretische Physik T39, Physik-Department, TU M{\"u}nchen, 
    D-85747 Garching, Germany}
  \\[2ex]
  \myaddress{$^b$ ECT*, Villa Tambosi, I-38050 Villazzano (Trento), Italy}
  \\[2ex]
   \myaddress{$^c$ Dipartimento di Fisica Nucleare e Teorica, Universit\'a 
    degli Studi di Pavia and INFN, Sezione di Pavia, Pavia, Italy}

%\vspace*{0.2cm}
\end{center}

%\vspace*{0.5cm}

\begin{abstract}
  We present a projector formalism which allows to define dynamical polarizabilities 
of the nucleon from a multipole expansion of the nucleon Compton amplitudes. We 
give predictions for the energy dependence of these dynamical polarizabilities both 
from dispersion theory and from leading-one-loop chiral effective field theory. Based 
on the good agreement between the two theoretical frameworks, we conclude that the 
energy dependence of the dynamical polarizabilities is dominated by chiral dynamics, 
except in those multipole channels where the first nucleon resonance $\Delta$(1232) 
can be excited. Both the dispersion theory framework and a chiral effective field theory 
with explicit $\Delta$(1232) degrees of freedom lead to a very good description of the
available low energy proton Compton data. We discuss the sensitivity of the proton 
Compton cross section to dynamical polarizabilities of different multipole content and 
present a fit of the static electric and magnetic dipole polarizabilities from low-energy 
Compton data up to $\omega \sim 170\;\mathrm{MeV}$, finding
$\bar{\alpha}_E=(11.04\pm1.36)\cdot 10^{-4}\;\mathrm{fm}^3,\;
 \bar{\beta}_M =( 2.76\mp1.36)\cdot 10^{-4}\;\mathrm{fm}^3$.
\end{abstract}

%\vskip 1.0cm

\noindent
\begin{tabular}{rl}
Suggested PACS numbers:& 11.55.Fv, 13.40.-f, 13.60.Fz, 14.20.Dh \\[1ex]
Suggested Keywords: &\begin{minipage}[t]{11cm}
                    Effective Field Theory, Nucleon Polarizabilities,\\
                    Compton scattering, Nucleon Structure, Dispersion
                    Relations. 
                    \end{minipage}
\end{tabular}

%\vskip 1.0cm

\end{titlepage}

\setcounter{page}{2} \setcounter{footnote}{0} \newpage

%%%%%%%%%%%%%%%%%%%%%%%%%%%%%%%%%%%%%%%%%%%%%%%%%%%%%%%%%%%%%%%%%%%%%%%%%%%%%%%
%%%%%%%%%%%%%%%%%%%%%%%%%%%%%%%%%%%%%%%%%%%%%%%%%%%%%%%%%%%%%%%%%%%%%%%%%%%%%%%
%%%%%%%%%%%%%%%%%%%%%%%%%%%%%%%%%%%%%%%%%%%%%%%%%%%%%%%%%%%%%%%%%%%%%%%%%%%%%%%
% Main Body
%

%%%%%%%%%%%%%%% Intro %%%%%%%%%%%%%%%%%%%
\section{Introduction}
\setcounter{equation}{0}
\label{sec:intro}

Compton scattering off a proton has a long history in the field of nucleon structure 
studies with electromagnetic probes \cite{Report}. While for photon energies below 25 MeV in the center-of-mass (cm) system the experimental cross section is well described with the assumption of a point-like spin 1/2 nucleon with an additional anomalous magnetic moment $\kappa$ \cite{Powell}, the internal structure of the nucleon starts to play a role at higher energies. Nowadays, these nucleon structure effects have been known for many decades and (in the case of a proton target) quite reliable theoretical calculations for the deviations from the Powell-cross section exist, typically parameterized in terms of the electromagnetic polarizabilities of the nucleon, of which the (static) electric and magnetic dipole polarizabilities $\bar{\alpha}_E,\,\bar{\beta}_M$ are only the most prominent examples \cite{Report}. While the general sizes of the dominant polarizabilities extracted from proton Compton experiments with the help of theoretical frameworks based on Dispersion Theory\footnote{For details on the various variants of Dispersion Theory we refer to the discussion given in Ref.~\cite{Report}.} have only received minor modifications over the past decade, several calculations were undertaken trying to identify the active constituent degrees of freedom within the nucleon, which are responsible for these structure effects. These theoretical calculations come from a wide set of theoretical frameworks, covering the range from constituent quark degrees of freedom ({\it e.g.} see \cite{CD97} and references therein) to the role of pionic fluctuations originating from chiral dynamics in the nucleon \cite{CD97,CD00}. 

In principle, nucleon Compton scattering can provide a wealth of information about the internal structure of the nucleon. However, in contrast to many other 
electromagnetic processes---{\it e.g.} pion photo-production off a nucleon---the nucleon structure effects probed in Compton scattering in most of the recent
analyses have not been analyzed in terms of a multipole expansion \cite{Pfeil}. Instead, most experiments have focused on just two structure parameters, which in analogy to classical electrodynamics are interpreted as the very (static) electric and magnetic polarizabilities $\bar{\alpha}_E$ and $\bar{\beta}_M$ mentioned above. Therefore, at present, quite different theoretical frameworks are able to provide a consistent, qualitative picture for the leading static polarizabilities $\bar{\alpha}_E,\,\bar{\beta}_M$ \cite{Report}.

In order to obtain a better filter for the theoretical mechanisms proposed for the internal structure of the nucleon as seen in nucleon Compton scattering, 
it was pointed out in Ref.~\cite{GH1} that the two concepts of Compton multipoles and nucleon polarizabilities can be combined if one introduces so-called ``dynamical polarizabilities" of the nucleon. These dynamical polarizabilities are functions of the excitation energy and encode the dispersive effects of $\pi N,\;N^\ast$ and other higher intermediate states \cite{GH1}. In the limit of zero excitation energy one regains the usual (static) polarizabilities $\bar{\alpha}_E,\,\bar{\beta}_M$. Extensions to higher multipole channels or (static) spin-polarizabilities discussed in the literature \cite{babusci} are straightforward and will be discussed in Sect.~2.

In this work, we go beyond the theoretical concept study of Ref.~\cite{GH1} and present a first analysis \cite{thesis,talks} of  the sensitivity of proton Compton cross section data to these dynamical polarizabilities. 
This paper is organized as follows: In Sect.~2, we present the basic formalism for a multipole expansion in nucleon Compton scattering and give the connection between the Compton multipoles and the concept of dynamical polarizabilities. In Sect.~3, we briefly discuss the two theoretical frameworks---Dispersion Theory and Chiral Effective Field Theory---employed here to study nucleon Compton scattering. In Sect.~4, we then confront the theoretical calculations with actual proton cross section data. In Sect.~5 we present a detailed analysis of the physics contained in the dynamical polarizabilities at energies below the $\Delta$(1232) resonance and in Sect.~6 we briefly cover the behaviour of dynamical polarizabilities in the resonance region as predicted by Dispersion Theory. Having relegated a lot of the necessary formulae to the Appendices A-D, we then come to the conclusions of this study.

%%%%%%%%%%%%%%% Intro %%%%%%%%%%%%%%%%%%%
\section{Multipole Expansion for Nucleon Compton Scattering}
\setcounter{equation}{0}
\label{sec:formalism}
%%%%%%%%%%%%%%%%%%%%%%%%%%%%

%%%%%%%%%%%%%%%%%%%%%%%%%%%%
\subsection{From Amplitudes to Multipoles}
\label{sec:multipoles}
%%%%%%%%%%%%%%%%%%%%%%%%%%%%

The $T$ matrix of real Compton scattering off the nucleon is written in terms
of six structure amplitudes $R_i(\omega,z),\;i=1,\dots,6$:
\begin{eqnarray}
T_{fi}&=&\frac{4\pi\,W}{M}\sum_{i=1}^6\rho_i\,R_i(\omega,z)
\label{Tmatrix}
\end{eqnarray}
For the Compton multipole expansion, we follow the tradition of Ritus et al. \cite{Ritus} and work in the centre-of-mass (cm) frame.  $W=\omega+\sqrt{M^2+\omega^2}$ is the total cm energy and $\omega$ denotes the cm energy of a real photon scattering under the cm angle $\theta$ off the nucleon with mass $M$ and $z=\cos\theta$. The basis functions $\rho_i$ read

\begin{eqnarray}
\rho_1=\vec{\epsilon}\,'^*\cdot \vec{\epsilon},\;\;\rho_2=\vec{s}\,'^*\cdot\vec{s},\;\;
\rho_3=i\,\vec{\sigma}\cdot\left(\vec{\epsilon}\,'^*\times \vec{\epsilon}\right),\;\;
\rho_4=i\,\vec{\sigma}\cdot\left(\vec{       s}\,'^*\times \vec{       s}\right),
\;\;\;\;\;\;\;\;\;\;\;\;\;\;\nonumber\\
\rho_5=i\,
 \left((\vec{\sigma}\cdot \hat{\vec{k}} ) \,(\vec{s       }\,'^*\cdot\vec{\epsilon})
      - (\vec{\sigma}\cdot \hat{\vec{k}}') \,(\vec{\epsilon}\,'^*\cdot\vec{s}       )\right),\;\;
\rho_6=i\,
 \left((\vec{\sigma}\cdot \hat{\vec{k}}') \,(\vec{s       }\,'^*\cdot\vec{\epsilon})-
       (\vec{\sigma}\cdot \hat{\vec{k}} ) \,(\vec{\epsilon}\,'^*\cdot\vec{s}       )\right)
\label{basis}
\end{eqnarray}
with $\vec{s}     =\hat{\vec{k}} \times\vec{\epsilon},\;\,
      \vec{s}\,'^*=\hat{\vec{k}}'\times\vec{\epsilon}\,'^*$
and $\vec{\sigma}$ the vector of the Pauli spin matrices. Furthermore, 
$\hat{\kv}=\kv/\omega$
  ($\hat{\kv}{}^\prime=\kv^\prime/\omega$) is the unit vector in the direction
  of the momentum of the incoming (outgoing) photon with polarization
  $\vec{\epsilon}$ ($\vec{\epsilon}\,'^*$). 

\begin{figure}[!htb]
\begin{center} 
  \includegraphics*[width=.65\textwidth]{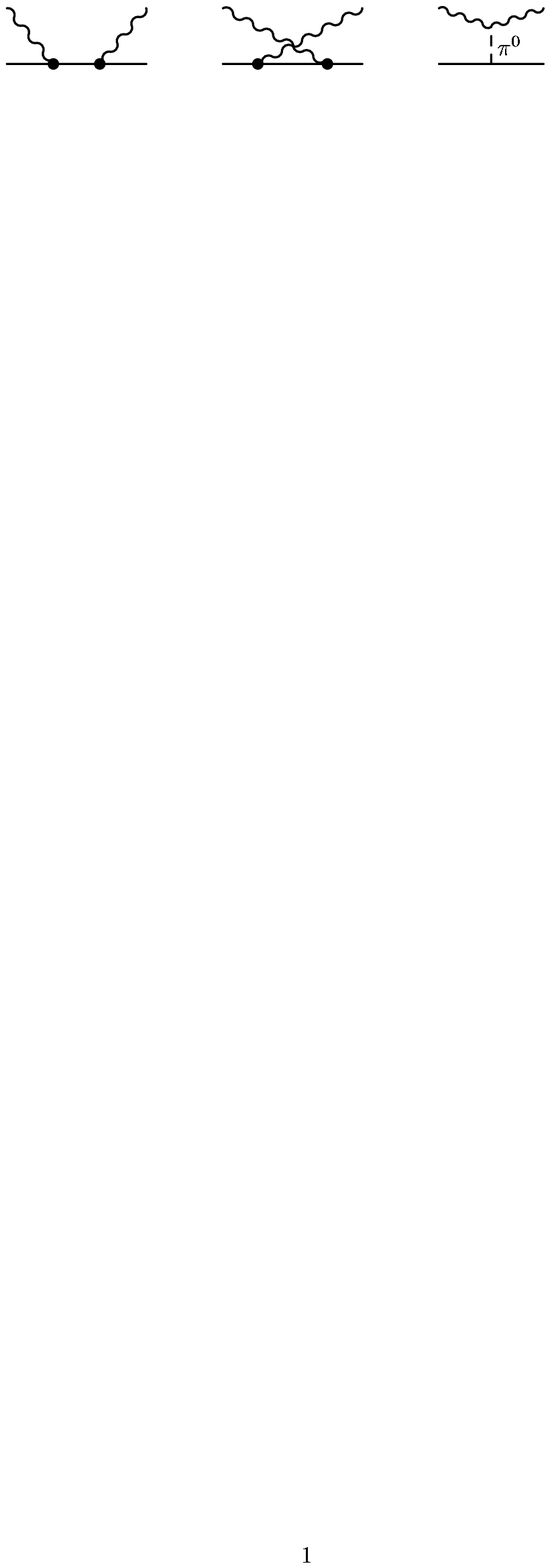}
\caption{Schematic representation of the three types of pole contributions to nucleon Compton scattering in the $s$-, $u$- and $t$-channel.}\label{pole}
\end{center}
\end{figure}

While the multipole expansion can in principle be defined for the entire Compton amplitude,  the nucleon structure effects as for example expressed in $\bar{\alpha}_E$ and $\bar{\beta}_M$ are typically defined as intermediate states which go beyond single nucleon contributions. Traditionally, this corresponds to subtracting from the full amplitudes the Powell amplitudes \cite{Powell} of Compton scattering on a pointlike nucleon of spin $\frac{1}{2}$ and anomalous magnetic moment $\kappa$. Therefore, we separate the six amplitudes into
structure independent (pole) and structure-dependent (non-pole) parts,
\begin{eqnarray}
R_i(\omega,z)&=&R_i^{\mathrm{pole}}(\omega,z)+\bar{R}_i(\omega,z)\;\;. 
\label{BnB}
\end{eqnarray}
Note that here we define the pole contributions as those terms which have a nucleon pole in the $s$- or $u$-channel and {\em in addition} as those terms which have a pion pole in the $t$-channel. Schematically, we show these three contributions in Fig.~\ref{pole}
and note that any theoretical framework utilized to calculate Compton scattering has to clearly separate these pole contributions before any information on static or dynamical polarizabilities can be obtained. Obviously, the proton Compton cross-sections are irreverent to this artificial separation of the 
amplitudes.

As the pole contributions to nucleon Compton scattering are known for many decades \cite{Low}, the main interest in Compton studies over the past few years has been focused on the non-pole contributions $\bar{R}_i$. In Ref.~\cite{GH1}, it has been suggested to apply the Compton multipole expansion only to these structure-dependent terms. In analogy to Ref.~\cite{Ritus}, one obtains
\begin{eqnarray}\label{ritus}
\bar{R}_1(\omega,z)&=&\sum_{l=1}^\infty\bigg[[(l +1)f_{EE}^{l+}(\omega)+l
f_{EE}^{l-} (\omega)](l P_l^{\prime}(z)+ P_{l-1}^{\prime\prime}(z))
 \non\\&&\phantom{\sum_{l=1}^\infty\bigg[}
  -[(l+1)f_{MM}^{l+}(\omega)+l f_{MM}^{l-} (\omega)  ]
  P_l^{\prime\prime}(z) \bigg],\\
\bar{R}_2(\omega,z)&=&\sum_{l=1}^\infty\bigg[[(l +1)f_{MM}^{l+} (\omega)+l
f_{MM}^{l-} (\omega)](l P_l^{\prime}(z)+ P_{l-1}^{\prime\prime}(z)) 
 \non\\&&\phantom{\sum_{l=1}^\infty\bigg[}
  -[(l+1)f_{EE}^{l+} (\omega)+l f_{EE}^{l-} (\omega)  ]P_l^{\prime\prime}(z)
  \bigg],\\
\bar{R}_3(\omega,z)&=&\sum_{l=1}^\infty\bigg[[f_{EE}^{l+} (\omega)-
f_{EE}^{l-} (\omega)](  P_{l-1}^{\prime\prime}(z)-l^2  P_l^{\prime}(z))
  -[  f_{MM}^{l+} (\omega)-  f_{MM}^{l-} (\omega)  ]P_l^{\prime\prime}(z)
 \non\\&&\phantom{\sum_{l=1}^\infty\bigg[}
  +2f_{EM}^{l+} (\omega) P_{l+1}^{\prime\prime}(z)
  -2f_{ME}^{l+} (\omega)P_l^{\prime\prime}(z) \bigg],\\
\bar{R}_4(\omega,z)&=&\sum_{l=1}^\infty\bigg[[f_{MM}^{l+} (\omega)
-  f_{MM}^{l-} (\omega)](  P_{l-1}^{\prime\prime}(z)-l^2  P_l^{\prime}(z))
  -[  f_{EE}^{l+} (\omega)-  f_{EE}^{l-} (\omega)  ]P_l^{\prime\prime}(z)
 \non\\&&\phantom{\sum_{l=1}^\infty\bigg[}
  +2f_{ME}^{l+} (\omega) P_{l+1}^{\prime\prime }
  -2f_{EM}^{l+} (\omega)P_l^{\prime\prime}(z) \bigg],
\end{eqnarray}
\begin{eqnarray}
\bar{R}_5(\omega,z)&=&\sum_{l=1}^\infty\bigg[[f_{EE}^{l+} (\omega)-
f_{EE}^{l-} (\omega)](l P_l^{\prime\prime}(z)+P_{l-1}^{\prime\prime\prime}(z))
  -[  f_{MM}^{l+} (\omega)-  f_{MM}^{l-} (\omega)  ] P_l^{\prime\prime\prime }
  \\ 
& &\phantom{\sum_{l=1}^\infty\bigg[}+ f_{EM}^{l+}
(\omega)[(3l+1)P_l^{\prime\prime}(z)+2P_{l-1}^{\prime\prime\prime}(z)]-
f_{ME}^{l+}
(\omega)[(l+1)P_{l+1}^{\prime\prime}(z)+2P_l^{\prime\prime\prime}(z)]
\bigg],\non\\
\bar{R}_6(\omega,z)&=&\sum_{l=1}^\infty\bigg[[f_{MM}^{l+} (\omega)-
f_{MM}^{l-} (\omega)](l P_l^{\prime\prime}(z)+P_{l-1}^{\prime\prime\prime}(z))
  -[  f_{EE}^{l+} (\omega)-  f_{EE}^{l-} (\omega)  ]
  P_l^{\prime\prime\prime}(z)\\&&\phantom{\sum_{l=1}^\infty\bigg[}
  + f_{ME}^{l+}
  (\omega)[(3l+1)P_l^{\prime\prime}(z)+2P_{l-1}^{\prime\prime\prime}(z)] 
 - f_{EM}^{l+}
 (\omega)[(l+1)P_{l+1}^{\prime\prime}(z)+2P_l^{\prime\prime\prime}(z)]
 \bigg].\non 
\end{eqnarray}
The prime denotes differentiation with respect to $z=\cos\theta$ in the cm
system, and $P_l(z)$ is the $l$th Legendre polynomial. The functions 
$f_{TT^\prime}^{l\pm}(\omega)$ are the Compton multipoles and
correspond to transitions $T l\rightarrow T^\prime l^\prime$, where
$T,T^\prime=E,M$ labels the coupling of the incoming or outgoing photon as
electric or magnetic. Here $l$ ($l^\prime=l\pm\{1,\,0\}$) denotes the angular momentum of the initial (final) photon, whereas the total angular momentum is $l\pm=j=l\pm\half$.
We note that mixed multipole amplitudes $T\not=T^\prime$ only occur in the spin-dependent 
amplitudes $\bar{R}_i,\,i=3,\ldots, 6$. 

Having defined purely structure-dependent Compton multipoles in the cm frame, we now move on to connect them to polarizabilities.

%%%%%%%%%%%%%%%%%%%%%%%%%%%%
\subsection{Dynamical and Static Polarizabilities}
\label{sec:dynpolas}
%%%%%%%%%%%%%%%%%%%%%%%%%%%%%%%%%%%%

In order to derive a consistent connection between the Compton multipoles $f_{TT^\prime},\;T,T^\prime=E,M$
and the polarizabilities of definite spin structure and multipolarity at a
certain energy, we recall the low energy behavior of the multipoles
in the cm frame \cite{Ritus}:
\begin{eqnarray}
  &\dis f_{TT^\prime}^{l\pm}(\omega)\sim\omega^{2l}\;\;,\;\;T=T^\prime \;\;&\\
  &\dis
  f_{TT^\prime}^{l+}(\omega)\sim\omega^{2l+1}\;\;,\;\;T\neq 
  T^\prime \;\;.&
\end{eqnarray}
With this information, dynamical spin-independent electric or magnetic dipole and quadrupole polarizabilities were defined as linear combinations of Compton multipoles in \cite{GH1}:
\begin{eqnarray}
\alpha _{E1}(\omega)&=&\left[2\,f_{EE}^{1+}(\omega)+f_{EE}^{1-}(\omega)\right]/{\omega^2}\nonumber\\
\beta  _{M1}(\omega)&=&\left[2\,f_{MM}^{1+}(\omega)+f_{MM}^{1-}(\omega)\right]/{\omega^2}\nonumber\\
\alpha _{E2}(\omega)&=& 36\left[3\,f_{EE}^{2+}(\omega)+2\,f_{EE}^{2-}(
\omega)\right]/{\omega^4}\nonumber\\
\beta  _{M2}(\omega)&=& 36\left[3\,f_{MM}^{2+}(\omega)+2\,f_{MM}^{2-}
(\omega)\right] /{\omega^4}\label{spinindiepolasdef}
\end{eqnarray}
We note that the normalization of these linear superpositions has been chosen in such a way that the usual (static) electric and magnetic polarizabilities of the nucleon typically discussed in Compton scattering can be recovered from the dynamical dipole polarizabilities via
\begin{eqnarray}
\bar{\alpha}_E=\lim_{\omega\rightarrow 0}\alpha_{E1}(\omega)\;, \,\quad\quad
\bar{\beta}_M=\lim_{\omega\rightarrow 0}\beta_{M1}(\omega) \;.
\end{eqnarray}
Likewise, the static electric and magnetic quadrupole polarizabilities $\bar{\alpha}_{E2},\,\bar{\beta}_{M2}$ discussed in Ref.~\cite{babusci} and Ref.~\cite{holstein} (see also App.~D) can be obtained as the zero energy limit of the corresponding dynamical quadrupole polarizabilities.

Extending Ref.~\cite{GH1}, we also introduce dynamical {\em spin-dependent} dipole polarizabilities via
\begin{equation} \label{eq:spinpolasdef}
\begin{array}{r@{\;=\;}c l@{\rightarrow}r}
\gamma _{E1E1}(\omega)&\dis \left[
f_{EE}^{1+}(\omega)-f_{EE}^{1-}(\omega)\right]/{\omega^3} &(E1& E1)\;,\\[1ex] 
\gamma _{M1M1}(\omega)&\dis \left[
f_{MM}^{1+}(\omega)-f_{MM}^{1-}(\omega)\right]/{\omega^3} &(M1& M1)\;,\\[1ex] 
\gamma _{E1M2}(\omega)&\dis 
  6\,f_{EM}^{1+}(\omega)/{\omega^3}&(E1& M2)\;,\\[1ex]
\gamma _{M1E2}(\omega)&\dis  
 6\,f_{ME}^{1+}(\omega)/{\omega^3}&(M1& E2)\;.
\end{array}
\end{equation}
In the limit of zero photon energy $\omega\rightarrow 0$, one again recovers the four static spin-polarizabilities $\bar{\gamma}_{E1E1}$, $\bar{\gamma}_{M1M1}$, $\bar{\gamma}_{E1M2}$, $\bar{\gamma}_{M1E2}$ of the nucleon:
\begin{eqnarray}
\bar{\gamma}_{TlT^\prime l^\prime}&=&\lim_{\omega\rightarrow 0}
      \gamma_{TlT^\prime l^\prime}(\omega) \; ,\;T,T'=E,M.
\end{eqnarray}
Here, these four static spin polarizabilities are written in the so called multipole-basis \cite{babusci}. The connection to the Ragusa-basis $\gamma_i,\,i=1,\ldots, 4$ \cite{Ragusa}, is discussed in \cite{Hem01}. We note that at present there exists little information on the spin-dependent nucleon polarizabilities. Only two linear combinations are constrained from experiments \cite{Report}, typically denoted as the forward $\gamma_0$ and the backward $\gamma_\pi$ spin-polarizabilities of the nucleon. Via the connection
\begin{eqnarray}
\gamma_0&=& -\bar{\gamma}_{E1E1}-\bar{\gamma}_{E1M2}-\bar{\gamma}_{M1M1}-\bar{\gamma}_{M1E2}\nonumber\\
\gamma_\pi&=& -\bar{\gamma}_{E1E1}-\bar{\gamma}_{E1M2}+\bar{\gamma}_{M1M1}+\bar{\gamma}_{M1E2} \label{gpi}
\end{eqnarray}
one realizes that in each case all four (dipole) spin polarizabilities are involved. 

While the static polarizabilities of the nucleon are real, we note that the dynamical polarizabilities become complex once the energy in the intermediate state is high enough to create an on-shell intermediate state, the first being the physical $\pi N$ intermediate state. Below the two pion threshold, the imaginary parts of the dynamical polarizabilities can be understood very easily. They are simply given by the well-known multipoles of single pion photo-production ({\it e.g.} see \cite{HDT}). One obtains
\begin{eqnarray}
\label{eq:im_dyn_pol} 
\begin{array}{l@{\hq\hq}l}
  \Im [ \alpha_{E1} (\omega)] =\frac{k_{\pi}}{\omega^2}\sum_c (2|E^{(c)}_{2-}|^2+
  |E^{(c)}_{0+}|^2)\;,&
  \Im [ \beta_{M1} (\omega)] = \frac{k_{\pi}}{\omega^2}\sum_c
  (2|M^{(c)}_{1+}|^2+ |M^{(c)}_{1-}|^2)\;,\\[1ex]
  \Im [ \alpha_{E2} (\omega)] = 36 \; \frac{k_{\pi}}{\omega^4} 
  \sum_c (3|E^{(c)}_{3-}|^2+ |E^{(c)}_{1+}|^2)\;,&
  \Im [ \beta_{M2} (\omega)] = 36 \; \frac{k_{\pi}}{\omega^4} 
  \sum_c (3|M^{(c)}_{2+}|^2+ |M^{(c)}_{2-}|^2 )\;,\\[1ex]
  \Im [ \gamma_{E1E1} (\omega)] = \frac{k_{\pi}}{\omega^3} 
  \sum_c (|E^{(c)}_{2-}|^2- |E^{(c)}_{0+}|^2 )\;,&
  \Im [ \gamma_{M1M1} (\omega)] = \frac{k_{\pi}}{\omega^3} 
  \sum_c (|M^{(c)}_{1+}|^2- |M^{(c)}_{1-}|^2 )\;,\\[1ex]
  \Im [ \gamma_{E1M2} (\omega)] = 6\;\frac{k_{\pi}}{\omega^3} 
  \sum_c \Re[E^{(c)}_{2-}(M^{(c)}_{2-})^* ]\;,&
  \Im [ \gamma_{M1E2} (\omega)] = -6\;\frac{k_{\pi}}{\omega^3} 
  \sum_c \Re [E^{(c)}_{1+}(M^{(c)}_{1+})^* ]\;,
\end{array}\nonumber\\
\end{eqnarray}
where $k_{\pi}$ is the magnitude of the pion momentum and $E_{l\pm}^c$ and
$M_{l\pm}^c$ are pion photo-production multipoles which are summed over the
different isotopic or charge channels $c$. In the following, we therefore focus on the real parts of the dynamical polarizabilities and treat the imaginary part as explained in Sect.~\ref{sec:dispersionformalism}.

This concludes our section pertaining to the definitions of the dynamical polarizabilities and their connection to static polarizabilities as well as to single pion photo-production. Before we discuss the numerical values of the (static) polarizabilities in the upcoming section, we first provide some background on the theoretical machinery employed to analyze nucleon Compton scattering.

%%%%%%%%%%%%%%% Intro %%%%%%%%%%%%%%%%%%%
\section{Theoretical Frameworks}
\setcounter{equation}{0}
\label{sec:dispersion}
%%%%%%%%%%%%%%%%%%%%%%%%%%%%

%%%%%%%%%%%%%%%%%%%%%%%%%%%%
\subsection{Dispersion Relation Analysis}
\label{sec:dispersionformalism}
%%%%%%%%%%%%%%%%%%%%%%%%%%%%%%

Unsubtracted fixed-$t$ dispersion relations (DR's) have already been applied
to a multipole analysis of Compton scattering in the $\Delta$-resonance
region~\cite{Huen97}, focusing on the possibility to combine simultaneously
the experimental information on pion-photoproduction and Compton scattering
multipoles for a study of the magnetic and electric $\Delta$-photoexcitation.
Because unsubtracted dispersion relations suffer from theoretical
uncertainties due to slow convergence of the integrals, an alternative scheme
has been proposed to consider subtracted DR at constant $t$~\cite{DGPV99}.  In
the following, we outline the essentials of the subtracted DR approach and in
particular we review the formalism to set up a multipole decomposition of the
Compton scattering amplitude which enters in the calculation of the dynamical
polarizabilities. We work with the set of invariant amplitudes
$A_i^{L}(\nu,t)$ introduced by L'vov et al.~\cite{Lvo97} which are functions
of $\nu=(s-u)/4M$ and $t=-2k\cdot k^\prime$, with $s,\;t$ and $u$ the
Mandelstam variables and $k^\mu$ ($k^{\mu\prime}$) the four-momentum of the
incoming (outgoing) photon. These amplitudes are free of kinematical
singularities and constraints, and satisfy the relations  
%except for crossing symmetry 
$A_i^L(\nu,t)=\;A_i^L(-\nu,t)$ due to crossing symmetry. 
Assuming furthermore analyticity and an appropriately soft
high-energy behavior, they fulfill the following fixed-$t$ DR's that are once
subtracted at $(\nu=0,t)$~\cite{DGPV99}:
\begin{eqnarray}
\mathrm{Re} [A_i^L(\nu, t)] \;&=&\; A_i^{L,\;\mathrm{N-pole}}(\nu, t) \;+\; 
\left[ A_i^L(0,t) -A_i^{L,\;\mathrm{N-pole}}(0, t) \right] \nonumber\\
&+&\;\frac{2}{ \pi}\;\nu^2\;  
\mathcal{P} \int_{\nu_0}^{+ \infty} d\nu' \; \frac{  
\mathrm{Im}_s[A_i^{L}(\nu',t)]}{\nu' \; (\nu'^2 - \nu^2)}\;. \label{eq:sub}
\end{eqnarray}
where $A_i^{L,\;\mathrm{N-pole}}(\nu, t)$ are the nucleon pole terms as
given in App.~A of Ref.~\cite{Lvo97}, and $\mathrm{Im}_s [A_i^L]$ the
discontinuities across the $s$-channel cuts of the Compton process which start
at the first inelastic threshold due to pion-nucleon intermediate states,
$\nu_0=m_\pi+(m_\pi^2+t/2)/(2M)$. The integration is understood as using the
principle value prescription, starting at the first threshold $\nu_0$.

Due to the subtraction at $\nu=0$, six subtraction functions $A_i^L(0,t)$
appear in Eq.~(\ref{eq:sub}) which are evaluated by setting up once-subtracted DR,
this time in the variable $t$:
\begin{eqnarray}
  \lefteqn{A_i^L(0, t) \;-\; A_i^{L,\;\mathrm{N-pole}}(0, t) =a_i \;+\;
\left[A_i^{L,\;\pi^0-\mathrm{pole}}(0, t) \;-\; 
      A_i^{L,\;\pi^0-\mathrm{pole}}(0, 0)\right] }\nonumber\\
&&+\;\frac{t}{\pi} \; \int_{(2 m_\pi)^2}^{+ \infty} dt' \; \frac{\mathrm{Im}_t
    [A_i^L(0,t')]}{t' \; (t' - t)} \;-\;\frac{t}{\pi} \; \int_{-
  \infty}^{-2 m_{\pi}^2 - 4 M m_\pi} dt' \;
  \frac{\mathrm{Im}_t [A_i^L(0,t')]}{t' \; (t' - t)} \;,
\label{eq:subt}
\end{eqnarray}
where $A_2^{L,\;\pi^0-\mathrm{pole}}(0, t)$ represents the contribution of the
$\pi^0$-pole in the $t$-channel, and the six coefficients $a_i\equiv
\;A_i^L(0, 0) \;-\; A_i^{L,\;\mathrm{N-pole}}(0, 0)$ are related to the static
polarizabilities as explained below.

In order to evaluate the dispersion integrals in Eq.~(\ref{eq:sub}), the 
imaginary parts in the $s$-channel are calculated from the unitarity relation,
taking into account the $\pi N$ intermediate states because of their strong
dominance in the kinematic regime we are interested in, $\nu \le 300 $ MeV. In
fact, due to the energy denominator $1/\left(\nu'(\nu'^2-\nu^2)\right)$ in the subtracted
dispersion integrals, the contributions from double-pion photoproduction and
other inelastic channels with thresholds at higher energies are largely
suppressed, and may be taken into account reliably by simple models.

In particular, we calculate the dominant $\gamma N\rightarrow \pi N
\rightarrow \gamma N$ contribution using the pion-photoproduction multipoles
of Hanstein et al.~\cite{HDT} up to energies of $\nu \approx 500 $ MeV.  At
higher energies (up to $\nu \simeq 1.5$ GeV), we take the recent SP02K
multipole solution of the SAID analysis~\cite{said}.  Of all multi-pion
intermediate states, we only take into account the resonance contribution, as
explained in Ref.~\cite{DGPV99}.

The subtracted $t$-channel integrals in Eq.~(\ref{eq:subt}) run along the
positive-$t$ channel cut from $4 m_\pi^2 \rightarrow + \infty$ and along the
negative-$t$ cut from $-\infty$ to $a=-2\, (m_{\pi}^2 + 2 M m_\pi) \approx -
0.56$~GeV$^2$.  At positive $t$, the $t$-channel discontinuities in $A_i^L$
can be evaluated by unitarity from the possible intermediate states for the
$t$-channel process $\gamma\gamma\rightarrow N\bar{N}$.  Since we only want to
evaluate the Compton amplitudes for small $t$, the subtracted dispersion
integrals are well saturated by the contribution of $\pi \pi$ intermediate
states.  As explained in detail in Ref.~\cite{DGPV99}, we calculate this
contribution by evaluating a unitarized amplitude for the $\gamma\gamma
\rightarrow \pi\pi$ subprocess and then combine it with the $\pi \pi
\rightarrow N\bar N$ amplitudes as determined in Ref.~\cite{Hoe83} from
dispersion theory by analytical continuation of $\pi N$ scattering amplitudes.
  The integral along the negative-$t$ cut is highly suppressed by the 
denominator of the subtracted DR for values of $|t|<<|a|$ and gives only a
small contribution. 
In order to take into account the main effects of the
negative-$t$ integrals, we evaluate the contributions of the $\Delta$-resonance
and non-resonant $\pi N$ intermediate states to the imaginary part of the 
Compton amplitudes and extrapolate these contributions into the 
%estimate the discontinuities $\mathrm{Im}_t
%[A_i^L(0,t)]$ by saturating with the $\Delta$-resonance.  In addition, we
%estimate the non-resonant $\pi N$ intermediate states by extrapolating the
%corresponding contribution to the $s$-channel Compton amplitudes into the
unphysical region at $\nu=0$ and negative $t$ by analytical continuation.

Finally, the relations between the six subtraction constants $a_i$ in
Eq.~(\ref{eq:subt}) and the static polarizabilities of Section~\ref{sec:dynpolas} are
\begin{eqnarray}
\bar{ \alpha}_E=-\frac{1}{4\pi}(a_1+a_3+a_6)\,,\quad\quad\bar{ \beta}_M=-\frac{1}{4\pi}(a_3+a_6-a_1)
\end{eqnarray}
for the spin-independent sector, and
\begin{alignat}{2}
   \bar{ \gamma}_{E1E1}&=\frac{1}{8\pi M}(a_6-a_4+2 a_5+a_2), &\qquad\qquad
  \bar{  \gamma}_{M1M1}&=\frac{1}{8\pi M}(a_6-a_4-2a_5-a_2),
  \nonumber\\
   \bar{ \gamma}_{E1M2}&=-\frac{1}{8\pi M}(a_4+a_6-a_2),&\qquad\qquad
   \bar{ \gamma}_{M1E2}&=-\frac{1}{8\pi M}(a_4+a_6+a_2) \nonumber
\end{alignat}
for the static spin-dependent polarizabilities.  In the Dispersion Theory calculation, we take as input the experimental values of the
electric and magnetic dipole polarizabilities as well as the backward spin
polarizability $\gamma_\pi$ of Eq.~(\ref{gpi}), fixing in this way the values of 
$a_1, a_2,$ and $a_3$. The remaining subtraction constants $a_4, a_5$, and $a_6$ are
calculated via the unsubtracted sum rules
\begin{equation}
  a_{i}=\frac{2}{\pi}\int_{\nu_0}^{+ \infty} d\nu' \; 
  \frac{\mathrm{Im}_s [A^L_{i}(\nu',t=0)]}{\nu'}\;. 
  \label{eq:unsub}
\end{equation}
 In particular, for the proton we use the following values from the recent global fit 
of Ref.~\cite{Olmos01}
\begin{equation}
 \bar{ \alpha}_E^{p}+\bar{\beta}_M^{p}=13.8 \times 10^{-4} \: \mbox{fm}^3\;\;,\;\;
 \bar{ \alpha}_E^{p}-\bar{\beta}_M^{p}=10.5 \times 10^{-4} \: \mbox{fm}^3\;\;,\;\;
\gamma_\pi^{p}=10.6 \times 10^{-4} \: \mbox{fm}^4\;\;,
  \label{eq:fit_pol_proton}
\end{equation}
where we have subtracted the contribution from the $\pi^0-$pole in
$\gamma_\pi^{p}$, $\gamma_\pi^{p,\: \pi^0-\mathrm{pole}}=-46.7 \times 10^{-4}
\: \mbox{fm}^4$. For the neutron, the static values of the lowest order
polarizabilities are fixed to
\begin{equation}
\bar{ \alpha}_E^{n}+\bar{ \beta}_M^{n}=15.2 \times 10^{-4} \: \mbox{fm}^3\;\;,\;\;
\bar{ \alpha}_E^{n}-\bar{ \beta}_M^{n}=9.8 \times 10^{-4} \: \mbox{fm}^3\;\;,\;\;
\gamma_\pi^{n}=13.6 \times 10^{-4} \: \mbox{fm}^4\;\;,
\label{eq:fit_pol_neutron}
\end{equation}
where the values for the sum and difference of the spin-independent neutron polarizabilities are taken from Ref.~\cite{Schumacher}, while the spin polarizability is calculated through fixed-$t$ unsubtracted DR's, as given in  Eq.~(\ref{eq:unsub}).
  
Once the dispersion results for the invariant amplitudes $A_i^L$ are obtained,
we calculate the helicity amplitudes through the relations given in
Eq.~(\ref{eq:hel_inv}) of App.~\ref{sec:AppA} and finally obtain the Compton multipoles $f_{TT^\prime}^{l\pm}(\omega)$ of Eq.~(\ref{ritus}) via the projection formulae of Eq.~(\ref{eq:phi_f}). We now move on to demonstrate how the same Compton multipoles can be obtained from Chiral Effective Field Theory ($\chi$EFT).

%%%%%%%%%%%%%%%%%%%%%%%%%%%%%%%%%%
\subsection{Chiral Effective Field Theory}
\label{subsec:hbchpt}
%%%%%%%%%%%%%%%%%%%%%%%%%%%%%%%%%%%%

Many calculations of nucleon Compton scattering---some even up to next-to-leading one loop order---have been performed in $\chi$EFT during the past decade \cite{CD97,CD00,BKM,latest}. Here, we extract information on the dynamical polarizabilities of the nucleon both from the leading-one-loop\footnote{We refrain from analyzing any HB$\chi$PT results beyond leading-one-loop order for nucleon Compton scattering at this point, because it appears that the concept of ${\cal O}(p^4)$ corrections used so far has to be modified \cite{Veronique}.} Heavy Baryon Chiral Perturbation Theory (HB$\chi$PT) calculation of Ref.~\cite{BKM} as well as from the leading-one-loop ``Small Scale Expansion" (SSE) calculations of Refs.~\cite{HHK97,HHKK}. We just note that HB$\chi$PT only involves explicit $\pi N$ degrees of freedom, whereas the SSE formalism is one possibility to also systematically include explicit spin 3/2 nucleon resonance degrees of freedom like $\Delta$(1232) in $\chi$EFT, and refer the interested reader to the literature \cite{BKM,review} for technical details.

\begin{figure}[!htb]
  \begin{center}
    \includegraphics*[width=.8\textwidth]{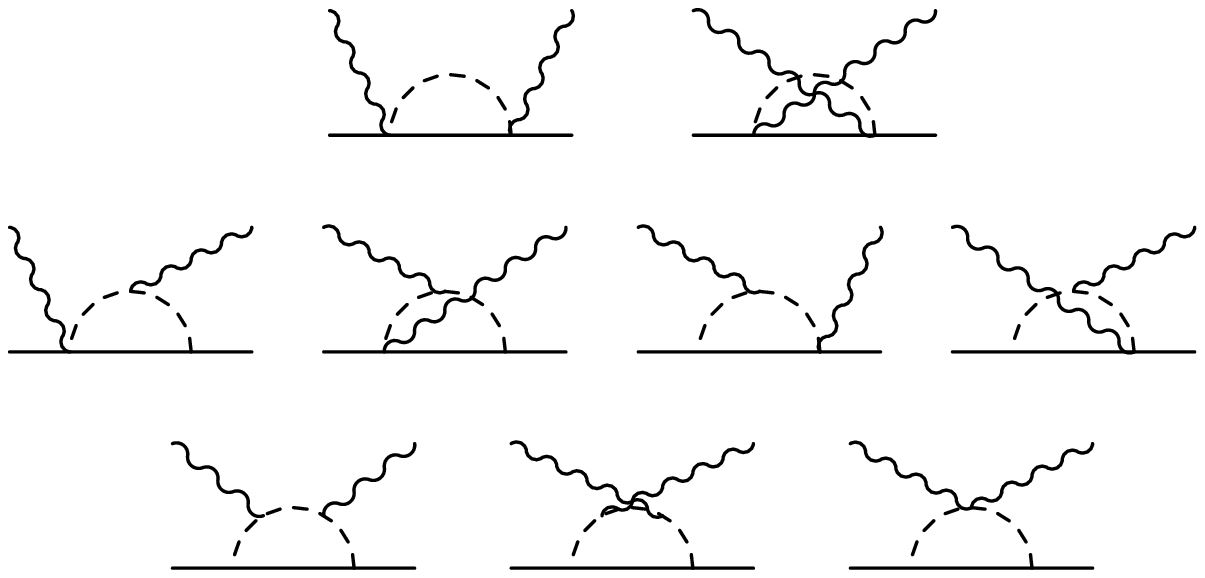}
    \caption{Leading-one-loop $N\pi$ continuum contributions to nucleon polarizabilities.}
    \label{Npicontinuum}
  \end{center}
\end{figure}

\begin{figure}[!htb]
  \begin{center}
    \includegraphics*[width=.8\textwidth]{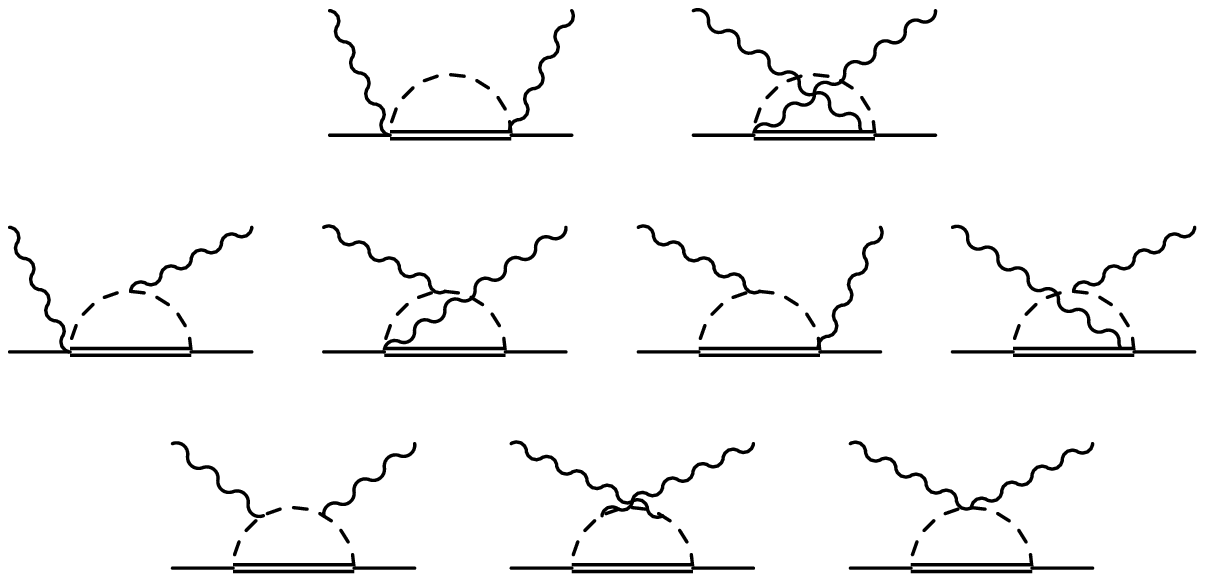}
    \caption{Leading-one-loop $\Delta\pi$ continuum contributions to nucleon polarizabilities.}\label{Deltapicontinuum}  
\end{center}
\end{figure}

\begin{figure}[!htb]
\begin{center}
  \includegraphics*[width=.65\textwidth]{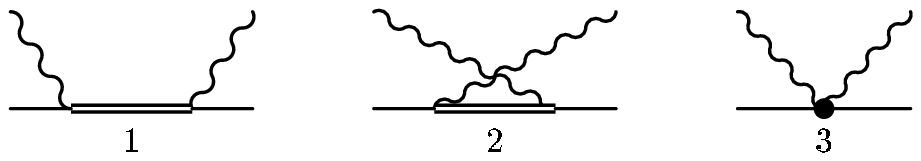}
\caption{$\Delta$ pole and short-distance contributions to nucleon polarizabilities} 
\label{Deltapolediagrams}
\end{center}
\end{figure}

The pole contributions to nucleon Compton scattering off the proton at leading-one-loop order in $\chi$EFT are given in Eq.~(\ref{eq:pole}) of App.~\ref{sec:AppB} for completeness. As discussed in the previous section, it is the non-pole contribution to Compton scattering which determines the polarizabilities.
In HB$\chi$PT, these structure-dependent contributions are solely given by $\pi N$ intermediate states (Fig.~\ref{Npicontinuum}), whereas in SSE one in addition has to take into account $\pi\Delta$ (Fig.~\ref{Deltapicontinuum}) as well as $\Delta$(1232) $s$- and $u$-channel pole contributions (Fig.~\ref{Deltapolediagrams}).

We note that we go beyond the existing leading-one-loop HB$\chi$PT/SSE calculations \cite{BKM,HHK97,HHKK} of nucleon Compton scattering in four aspects:
\begin{itemize}
\item[1)] Both HB$\chi$PT and SSE are non-relativistic frameworks leading to a $1/M$ expansion of the amplitudes, where $M$ corresponds to the mass of the (nucleon) target. In the leading-one-loop structure amplitudes $\bar{R}_i$ the one-pion production threshold
\begin{eqnarray}
\omega_\pi&=&\frac{m_\pi^2+2m_\pi M}{2(m_\pi+M)}\approx 131\;\MeV
\end{eqnarray}
is therefore not at the correct location. We correct for this purely kinematical effect by replacing the photon energy $\omega$ with the Mandelstam variable $s$ via
\begin{equation}
\omega\rightarrow\sqrt{s(\omega)}-M\; .\label{substitut}
\end{equation}
Obviously this replacement should only be applied in those places where an imaginary part arises above threshold (for further details we refer the reader to App. \ref{sec:AppB}). We implement this kinematical correction in the leading-one-loop $\pi N$-continuum contribution to the $\chi$EFT amplitudes throughout this work. The amplitudes thus modified are shown explicitly in App.~\ref{sec:AppB} for the case of SSE. We note that, strictly speaking, such kinematical corrections should be employed in non-relativistic $\chi$EFTs at all particle thresholds. However, given that the $\chi$EFTs employed here are valid only below the $\Delta$(1232) resonance, the one-pion production threshold is the only one to be taken care of.

\item[2)] Another kinematical effect concerns the exact location of the $\Delta$(1232) pole. In Ref.~\cite{HDT} it was determined as a $T$-matrix pole in the complex $W=\sqrt{s}$ plane at the location $M_\Delta=(1210-\,i\,50)$~MeV. We therefore employ the same substitution prescription for $\omega$ as in Eq.~(\ref{substitut}) in $s$-channel pole contributions of the $\Delta$(1232) resonance. Given that $\Delta$(1232) pole contributions in the $u$-channel can also affect higher multipoles, we make an analogous replacement $\omega\rightarrow M-\sqrt{u}$ in the $\Delta$(1232) $u$-channel pole contributions. While these kinematical details are of minor importance when one only discusses static polarizabilities (with the exception of $\bar{\beta}_{M2}$, see App.~\ref{sec:quadrupole}), they do become important in dynamical polarizabilities, when the photon energy is higher than 100 MeV. We note again that via these modifications, we have not introduced any additional physics content into the $\chi$EFT calculations, as in the $M\rightarrow\infty$ limit all these purely kinematical modifications reduce to the strict ${\cal O}(\epsilon^3)$ truncation of SSE \cite{HHK97,HHKK}. The detailed form of the modified amplitudes can be found in App.~\ref{sec:AppB}.

\item[3)] The parameters required for the leading-one-loop HB$\chi$PT calculation are well-known. For completeness, we list them in Table \ref{table1}. 
Also shown are the two parameters $\Delta_0$ and $g_{\pi N\Delta}$ utilized in the leading-one-loop SSE Compton scattering calculation of Refs.~\cite{HHK97,HHKK}. The numbers given here differ slightly from Ref.~\cite{HHK97}, as we determine them now from the exact kinematical location of the $\Delta$(1232)-pole in the complex $W$-plane, discussed in the previous paragraph. \\
To leading-one-loop order, the HB$\chi$PT calculation for nucleon Compton scattering is therefore parameter-free (in the sense that all parameters shown in Table \ref{table1} can be determined from sources outside Compton scattering). On the other hand, in the corresponding SSE calculation we are left with one free parameter $b_1$---which in $\chi$EFT corresponds to the leading $\gamma N\Delta$ coupling \cite{HHK97,review}. In Ref.~\cite{HHKK}, $b_1$ was estimated from the ``measured'' $\Delta\rightarrow N\gamma$ decay width to be $|b_1|\approx 3.9$. As this determination is very sensitive to the numerical value of the parameter $\Delta_0$ (for the value $\Delta_0=271$ MeV shown in Table \ref{table1}, we would obtain $|b_1|\approx 4.4$), we choose a different strategy here and determine $b_1$ directly from a fit to Compton cross section data.    

\item[4)] With the $\gamma N\Delta$ coupling constant $b_1$ as a fit parameter in the SSE analysis, we can constrain the crucial paramagnetic contribution from the $\Delta$ directly from data. However, it has been known for a long time that there must also be substantial diamagnetism in the nucleon---otherwise the small numbers for the static magnetic polarizability of the proton cannot be understood, see {\it e.g.} Ref.~\cite{Report} for details. At leading-one-loop order neither HB$\chi$PT nor SSE in their respective counting schemes allow for such a contribution \cite{HHK97}. Both schemes assume that this is a ``small'' higher order effect, which can be accounted  for at the next-to-leading one loop order. As a side remark we remind the reader that in Ref.~\cite{BKMS} it was shown in a next-to-leading one loop HB$\chi$PT calculation that for ``reasonable'' values of the regularization scale $\lambda$, a large part of this diamagnetism could be accounted for by $\pi N$ loop effects. Working only to leading-one-loop order, we cannot contribute to the discussion on the physical nature of this diamagnetism in the nucleon.

As we determine our paramagnetism from data and as at small photon energy there is this known delicate interplay between para- and diamagnetic contributions, we introduce two additional higher order $\gamma\gamma NN$ couplings $g_{117},\,g_{118}$ \cite{Fettes}
\begin{eqnarray}
\mathcal{L}_1^{CT}&=&\frac{g_{117}}{(4\pi \,f_\pi)^2\,M}
 \,\bar{N}\,v^\mu \,v^\nu \left<f_{\lambda\mu}^R
   \,f^{R\lambda}_\nu+f_{\lambda\mu}^L \,f^{L\lambda}_\nu\right>N,
\label{LCT1} \\
\mathcal{L}_2^{CT}&=&\frac{g_{118}}{(4\pi \,f_\pi)^2\,M}
 \,\bar{N} \left<f_{\mu\nu}^R\,f^{R\mu\nu}+
   f_{\mu\nu}^L\,f^{L\mu\nu}\right>N
\label{LCT2}
\end{eqnarray}
in the leading-one-loop SSE analysis, where $f^{\mu\nu}_R=f^{\mu\nu}_L
=\frac{e}{2}\,\tau_3\,\left(\partial^\mu A^\nu-\partial^\nu A^\mu\right)$ denote electromagnetic field tensors \cite{BKM}. Two independent structures are needed to separate magnetic and electric contributions via different linear combinations of $g_{117}$ and $g_{118}$. To promote these two structures to leading-one-loop order modifies the power counting, as they are  
formally part of the well-known next-to-leading one loop order chiral Lagrangean \cite{Fettes}.
In light of the reasoning given above, we nevertheless include them as free parameters in our SSE fit to Compton cross sections. If they turned out to give only small corrections, we could safely neglect them as a small higher order effect in accordance with the counting assumptions of SSE. However, as will be demonstrated in Sect.~\ref{subsec:fit}, this is not the case and these two couplings really have to be included already at leading-one-loop order, breaking the naive power-counting due to their unnaturally large sizes.
%The justification for it will follow in Sect.~\ref{sec:fit}, where we show that $g_{117}$, 
%$g_{118}$ do not obey the naive counting rules, as they turn out to be anomalously large. 
We find that the two couplings in Eqs.~(\ref{LCT1}, \ref{LCT2}) are sufficient to parameterize any unknown magnetic short-distance physics in nucleon Compton scattering (cf. Fig.~\ref{fig3}). 
%Formally, the two couplings of Eqs.~(\ref{LCT1}, \ref{LCT2}) are part of the well-known next 
%to leading-one-loop order chiral Lagrangean \cite{Fettes}. 
 The contributions of $g_{117},\,g_{118}$ to the Compton structure amplitudes are given explicitly in App. \ref{sec:AppB}.  
\end{itemize}
The leading-one-loop structure-dependent 
Compton amplitudes given in App.~\ref{sec:AppB} include the four modifications 
discussed above. In order to extract from them the dynamical polarizabilities of the 
nucleon in $\chi$EFT 
frameworks, one first projects out the Compton multipoles $f_{TT'}(\omega)$ of 
Sect.~\ref{sec:multipoles}, using the formulae in App.~\ref{sec:AppC}. The dynamical 
polarizabilities at definite multipolarity as a function of the photon energy follow then from 
Eqs.~(\ref{spinindiepolasdef}, \ref{eq:spinpolasdef}).

%When given such amplitudes 
%in $\chi$EFT, one first utilizes the projection formulae of App.~\ref{sec:AppB} and in such a 
%way determines the Compton multipoles $f_{TT^\prime}(\omega)$ discussed in 
%Section~\ref{sec:multipoles}. Finally, with help of Eq.~(\ref{spinindiepolasdef},
%\ref{eq:spinpolasdef}), one then arrives at the sought after expressions for dynamical 
%polarizabilities of the nucleon in $\chi$EFT frameworks.

This concludes our brief summary of leading-one-loop $\chi$EFT calculations for nucleon Compton scattering. We now move on to a determination of the three free parameters $b_1,\,g_{117}$ and $g_{118}$ from cross section data. 

\begin{table}[!htb] 
\begin{center}
\begin{tabular}{|c|c|c|}
\hline 
Parameter & Value & Comment \\
\hline 
$m_\pi$  & $139.6$ MeV & charged pion mass \\
$M$ & $938.9$ MeV & isoscalar nucleon mass \\
$f_\pi$  & $92.4$ MeV & pion decay constant \\
$g_A$ & $1.267$& axial coupling constant \\
$\alpha$ & $1/137$& QED fine structure constant \\
$\kappa_v$ & $3.71$ & isovector anom.~mag.~moment  \\
$\kappa_s$ & $-0.120$ & isoscalar anom.~mag.~moment \\
\hline
\rule{0ex}{2.5ex}
$\Delta_0$  & $271.1$ MeV & $N\Delta$ mass splitting \\
$g_{\pi N\Delta}$& $1.125$& $\pi N\Delta$ coupling constant \\
\hline
\end{tabular}
\end{center}
\caption{$\chi$EFT parameters determined independent of Compton scattering. Magnetic moments are given in nuclear magnetons.}\label{table1}
\end{table}

%%%%%%%%%%%%%%% Intro %%%%%%%%%%%%%%%%%%%
\section{Compton Cross Sections}
\setcounter{equation}{0}
\label{sec:cross}
%%%%%%%%%%%%%%%%%%%%%%%%%%%%

\subsection{General Remarks}

In the previous section, we have briefly reminded the reader of two theoretical frameworks which we now confront with actual Compton scattering data off a proton target.  This will also serve as a check for the parameters employed (in the case of Dispersion Theory and HB$\chi$PT), respectively provide us with the possibility to constrain some parameters (in the case of SSE). To be precise, we compare the experimental differential cross sections with predictions from Dispersion Theory, where the static values of the polarizabilities are fixed as described in Eqs.~(3.3)--(\ref{eq:fit_pol_neutron}), and with predictions from leading-one-loop HB$\chi$PT, which does not contain any free parameters to be determined from Compton scattering. In the case of leading-one-loop order SSE, we perform a fit of the three free parameters $b_1,\,g_{117},\,g_{118}$ discussed in the previous section to proton Compton data. In this section, we can therefore only check whether the three theoretical curves are consistent with the data. A detailed discussion of the electromagnetic structure of the proton in the three frameworks will be given in Sect~\ref{sec:comparison}.

So far, only spin-averaged cross sections on the proton have been measured. We
draw from a large set of data \cite{Olmos01,Hallin,Fed91,Mac95} covering proton Compton scattering from low energies to above the pion production
threshold. We present the low-energy data as functions of the differential cross section in the cm system versus the photon energy (in the cm system) at different angles $\theta_\mathrm{lab}$. Note that in the plots -- except for those, where we compare to the SAL data, which are given in the cm system -- we keep the scattering angle in the lab system because the data are given in this system. We also note that there are small deviations in the angles the various data sets are taken at, as described in the caption of Fig.~\ref{Olmos}.
%which we indicate via angular ranges in Fig.~\ref{Olmos} and Fig.~\ref{fig9}.
%but we neglect the minor corrections necessary: The data of \cite{Fed91} are not given at $59^\circ$ and $133^\circ$ but at $60^\circ$ and $135^\circ$; the data of~\cite{Mac95} are not given at $85^\circ$ and $133^\circ$ but at $90^\circ$ and $135^\circ$.

In the differential Compton scattering cross sections, the artificial
separation between pole and non-pole contributions is absent. Pole and non-pole contributions have to be added both in Dispersion Theory and in $\chi$EFT.
The differences between lab and cm system are expressed via the flux factors
\begin{equation}
  \Phi_\mathrm{cm} =\frac{M}{4\pi\,\sqrt{s(\omega)}}\;\;,\;\;
  \Phi_\mathrm{Lab}=\frac{\omega'_\mathrm{lab}}{4\pi\,\omega_\mathrm{lab}}
  \;\;, 
\label{phasespace}
\end{equation}
where $\omega'_\mathrm{lab}$ ($\omega_\mathrm{lab}$)
denotes the energy of the outgoing (incoming) photon in the lab frame.
In the spin-averaged case, the differential cross section is then given by
\begin{equation}
  \left.\frac{d\sigma}{d\Omega}\right|_\mathrm{frame}=\Phi_\mathrm{frame}^2
  \; |T|^2
\label{eq:diffcrosssection}
\end{equation}
with the absolute square of the Compton amplitude (see App.~\ref{sec:AppB} and Ref.~\cite{BKM})
\begin{equation}
\begin{split}
  |T|^2=&\;\frac{1}{2}\,|A^H_1|^2\,\left(1+z^2\right)+\frac{1}{2}\,|A^H_3|^2\,
  \left(3-z^2\right)\\
  &+\left(1-z^2\right)\left[4\,\Re[A^{H\dagger}_3\,A^H_6]+
    \Re[A_3^{H\dagger}\,A^H_4+2\,A_3^{H\dagger}\,A^H_5-A_1^{H\dagger}\,A^H_2
    ]\,z\right]\\
  &+\left(1-z^2\right)\biggl[\frac{1}{2}\,|A^H_2|^2\,\left(1-z^2\right)+
  \frac{1}{2}\,|A^H_4|^2\,\left(1+z^2\right)\\
  &+|A^H_5|^2\left(1+2z^2\right)+3|A^H_6|^2+
  2\Re[A_6^{H\dagger}\left(A^H_4+3A^H_5\right)]z+2\Re[A_4^{H\dagger} A^H_5]z^2
  \biggl]\;\;.\\
\end{split}
\end{equation}
After these general remarks, we now move on to the comparison with experiment.

%%%%%%%%%%%%%%%%%%%%%%%%%%
\subsection{Comparison to Experiment}
\label{subsec:experiment}
%%%%%%%%%%%%%%%%%%%%%%%%%%%

\begin{figure}[!htb]
  \begin{center}
    \includegraphics*[width=0.99\textwidth]{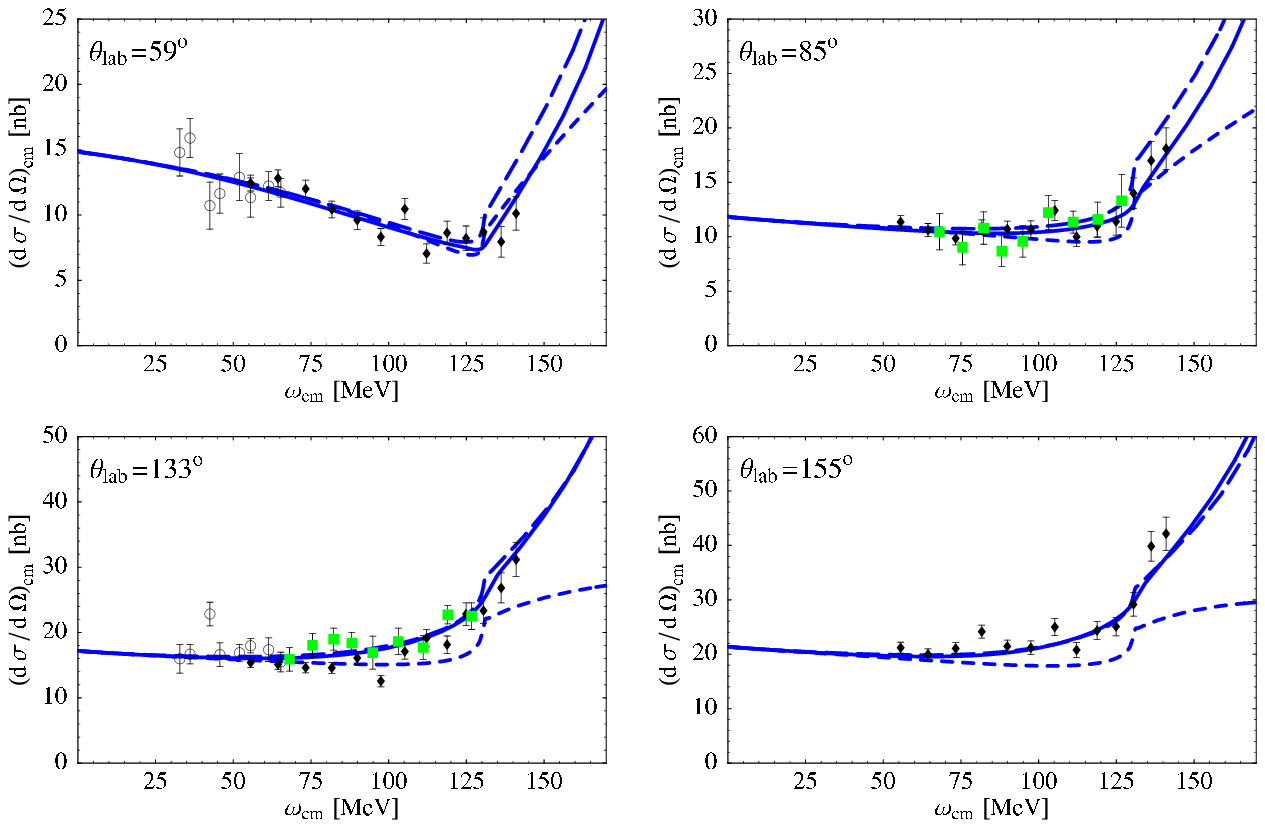}
 \caption{Comparison of the differential cross section data for
   Compton scattering off the proton (diamonds: Olmos de Leon et
   al.~\cite{Olmos01}, circles: Federspiel et al.~\cite{Fed91}, boxes:
   MacGibbon et al.~\cite{Mac95}) with Dispersion Theory and leading-one-loop order
 HB$\chi$PT respectively SSE at fixed lab angle. Solid line:
   DR results; short-dashed line: HB$\chi$PT; long-dashed line: SSE. Note that the 
data of~\cite{Fed91} are not given at $59^\circ$ and $133^\circ$ but at $60^\circ$ and 
$135^\circ$; the data of~\cite{Mac95} are not given at $85^\circ$ and $133^\circ$  but 
at $90^\circ$ and $135^\circ$.  }
  \label{Olmos}
  \end{center}
\end{figure}

\begin{figure}[!htb]
  \begin{center}
    \includegraphics*[width=0.99\textwidth]{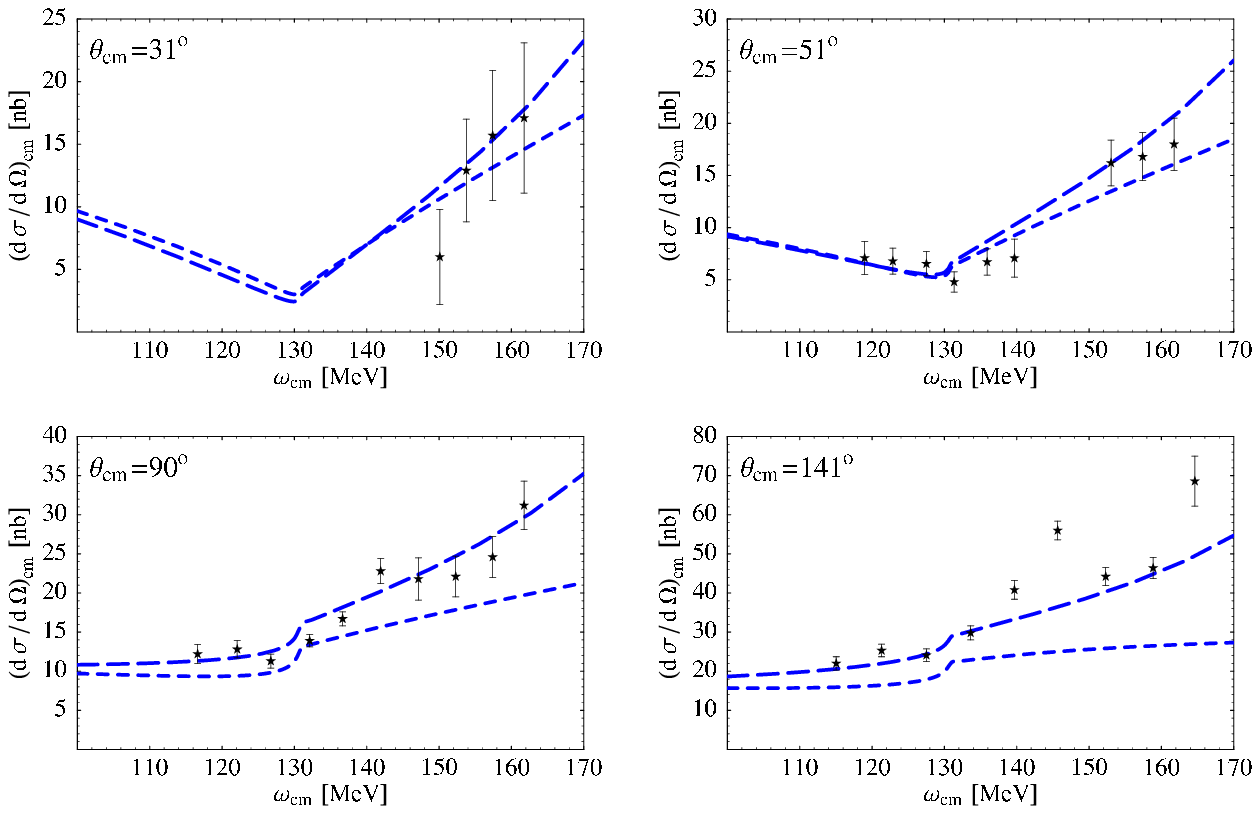}
 \caption{Comparison of the differential cross section data for
   Compton scattering off the proton from Hallin et al. \cite{Hallin} 
  with  leading-one-loop order
  HB$\chi$PT respectively SSE at fixed cm angle. 
  Short-dashed: HB$\chi$PT; long-dashed: SSE.}\label{Hallin}
  \end{center}
\end{figure}

Figs.~\ref{Olmos} and \ref{Hallin} compare several different cross section data at selected angles with the DR and HB$\chi$PT predictions, and with 
the result of our new SSE fit (details of the fit will be discussed in the next section). The data of Hallin et al. \cite{Hallin} (Fig.~\ref{Hallin}) provide important constraints for the fit above pion threshold. 
All three theoretical curves describe the available data quite well in the forward direction.  The upwards trend in the data above 130 MeV related to the opening of the single pion channel is also present in all three frameworks. However, while the SSE and DR results are rather similar at all angles, the HB$\chi$PT curve deviates from the data significantly in the backward direction, starting from photon energies around 80 MeV. The detailed analysis of the dynamical polarizabilities in the next section will show that this different energy dependence is due to the lack of explicit $\Delta$(1232) resonance degrees of freedom in HB$\chi$PT.
We find that cross section calculations in $\chi$EFT discarding the
$\Delta$ as explicit degree of freedom will always fail for large-angle 
scattering $\theta>90^\circ$, even at energies well below pion threshold. 

\begin{figure}[!htb]
  \begin{center}
    \includegraphics*[width=0.99\textwidth]{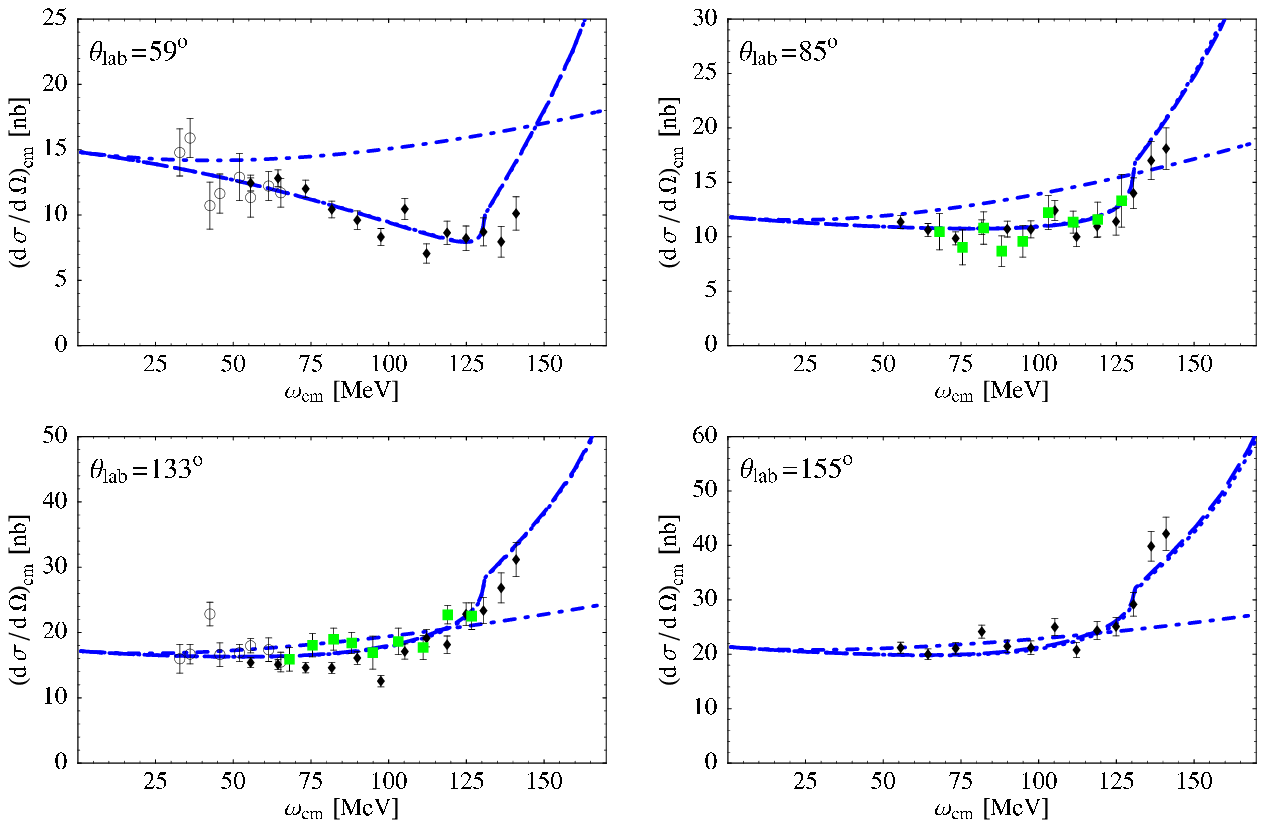}
 \caption{Comparison of the SSE multipole expansion to the differential cross section data for
   Compton scattering off the proton. Note that the $l=1$ and $l=2$ 
   truncations are 
   indistinguishable in 
   the energy region shown here. ($l=0$ truncation: dash-dotted curve; $l=1$ truncation: 
   dashed curve; $l=2$ truncation: dotted curve) }\label{fig9}
  \end{center}
\end{figure}

Having shown that the full Compton amplitudes $R_1,\ldots, R_6$ of DR and leading-one-loop SSE provide a good description of the available Compton data up to energies above pion threshold, we now determine what kind of physics dominates in the kinematic regime considered here. A well-established procedure to answer this question is of course a systematic multipole expansion of the Compton amplitudes $R_i(\omega,z)$ as discussed in Sect.~\ref{sec:multipoles}. In Fig.~\ref{fig9}, we compare the contributions of the first three terms of
the Compton multipole expansion to the same data as shown in Fig.~\ref{Olmos}. The curves plotted are for SSE, but the same pattern arises in DR. The $l=0$ truncation only contains the pole contributions to nucleon Compton scattering as shown by the diagrams in Fig.~\ref{pole}. Truncating the multipole expansion at $l=1$, the curve in addition 
contains all dynamical dipole polarizabilities.  All dynamical quadrupole
polarizabilities are contained in the $l=2$ truncation. As has been known for a
long time, a theoretical framework which only contains the pole contributions for nucleon Compton scattering gives a rather poor description of the cross sections, especially at small angles. The discrepancy between the $l=0$ result and the data is a clear indication of internal nucleon structure not contained in the standard pole terms. According to $\chi$EFT calculations, this structure can be interpreted as chiral dynamics in the nucleon: It is largely the contributions from the pions as the Goldstone Bosons of low energy QCD---or in other words the contribution from the pion cloud of the nucleon---which closes the gap between the pole contribution and the Compton data, at least for energies below the pion threshold. While this is not a surprise anymore after many years of $\chi$EFT calculations in nucleon Compton scattering, 
the surprising find from our multipole analysis is that up to energies of
$\omega\approx 200\;\MeV$, {\em there is no visible difference between the
$l=1$ and the $l=2$ truncation}! Therefore, the multipole expansion for nucleon Compton scattering converges very fast in this entire energy region. Furthermore, we see that aside from the well-known standard pole terms of Fig.~\ref{pole}, all one needs to know for a good description of nucleon Compton scattering are the six dynamical dipole polarizabilities $\alpha_{E1}(\omega),\,\beta_{M1}(\omega),\,\gamma_{E1E1}(\omega),\,\gamma_{M1M1}(\omega),\,\gamma_{E1M2}(\omega)$ and $\gamma_{M1E2}(\omega)$---as these 6 dynamical structures contain all the $l=1$ information. While $\chi$EFT calculations for nucleon Compton scattering in the past have either focused on the static values of the polarizabilities or on the (rather complicated) full Compton amplitudes $R_1,\ldots, R_6$, one can now dissect the role of chiral dynamics (and of explicit resonance contributions) in this process by looking at the individual multipole channels. 

Before we move on to a detailed comparison of DR and leading-one-loop order HB$\chi$PT, respectively SSE results for these six dynamical polarizabilities, we first have to give a few details regarding the three free parameters of SSE fitted to the Compton data.

%%%%%%%%%%%%%%%%%%%%%%%%%%%%%%%%%%%%%%
\subsection{SSE Fit and Static Dipole Polarizabilities}
\label{subsec:fit}
%%%%%%%%%%%%%%%%%%%%%%%%%%%%%%%%%%%%%
\subsubsection{Fit}
\label{subsubsec:fit}

The two short-distance terms containing the couplings $g_{117}$ and $g_{118}$ of Eqs.~(\ref{LCT1}, \ref{LCT2}) give contributions only to the electric and magnetic dipole polarizabilities and are energy-independent. The three free parameters of the leading-one-loop SSE analysis therefore correspond to a fit, which determines $\bar{\alpha}_E,\,\bar{\beta}_M$ plus the leading $\gamma N\Delta$ coupling $b_1$. For the fit, we use the data from \cite{Olmos01, Hallin}, and we show the results in Table~\ref{table2} together with their corresponding $\chi^2/d.o.f.$-values, which we calculated using the standard definition of
$\chi^2$, i.e. 
\begin{equation}
\chi^2=\sum\left(\frac{\sigma_\mathrm{exp}-\sigma_\mathrm{theo}}{\Delta\sigma}\right)^2
\label{eq:chisquared}
\end{equation}
with $\sigma_\mathrm{exp}$ the experimental, $\sigma_\mathrm{theo}$ the calculated 
cross-sections and $\Delta\sigma$ the experimental error bars. The number of degrees of 
freedom ($d.o.f.$) is the number of data points (115) minus the number of free 
parameters (3).
%We note that we have also analyzed the effect of the Baldin-constraint of 
%Eq.~(\ref{eq:fit_pol_proton}) for the sum of $\bar{\alpha}_E$ and $\bar{\beta}_M$, which reduces 
%the number of fit degrees of freedom by one. Putting in the Baldin constraint one obtains an 
%improved fit result for the static polarizabilities with smaller error bars, as expected. In 
%conclusion we can say that the obtained values for $\bar{\alpha}_E$ and $\bar{\beta}_M$ 
%compare very well with state of the art dispersion analyses \cite{Report}. 
Note that the value of $\bar{\alpha}_E+\bar{\beta}_M$ from the three-parameter-fit is consistent within error bars with the Baldin sum rule for the proton, 
$\bar{\alpha}_E+\bar{\beta}_M=13.8\cdot10^{-4}\;\mathrm{fm}^3$. One can therefore use the value of the Baldin sum rule as additional data point and reduce the number of free parameters to two. The results thus obtained are the ones we use in all our plots. The resulting static spin-independent dipole polarizabilities compare very well with state-of-the-art dispersion analyses \cite{Report}. Nevertheless, the $\chi^2/d.o.f.$-values of our fits are relatively large, but they are more an indication of the spread in the Compton data, which we have not allowed to float with a free normalization constant. The encouraging results of Table~\ref{table2} 
therefore prove, that by utilizing the SSE-amplitudes of App.~\ref{sec:AppB}, one has an alternative technique to extract the static polarizabilities $\bar{\alpha}_E,\;\bar{\beta}_M$ from low energy Compton data below the $\Delta$-resonance. We note that a determination of 
$\bar{\alpha}_E,\;\bar{\beta}_M$ from Compton data utilizing next-to-leading one loop order HB$\chi$PT has been presented in \cite{BeaneMcG}. The obtained results there are comparable to ours, although the 
authors had to restrict their fit to an angle-dependent $\omega_\mathrm{max}$, due to the known inadequate description of the Compton cross sections in backward direction in HB$\chi$PT.

We further note that the value we obtain
for the leading $\gamma N\Delta$ coupling agrees with previous analyses from the 
radiative $\Delta$-width as discussed in Sect.~\ref{subsec:hbchpt}, and, as a side 
remark, that 
we could also employ the strategy to rely on the DR-results for $\bar{\alpha}_E,\;
\bar{\beta}_M,\;\bar{\gamma}_{M1M1}$ to determine the three unknowns. In this case, the whole energy-dependence is predicted. The values thus obtained are identical with the fit-results within 
the error bars.

As our leading-one-loop order SSE-calculation only describes an isoscalar nucleon, we cannot contribute to the ongoing controversies over the size of the neutron polarizabilities \cite{Report,BeaneMcG,hgrieRupak}.

\begin{table}[!htb]
\begin{center}
\begin{tabular}{|c||r|r|r|}
\hline
Quantity&3-parameter-fit&2-parameter-fit&\cite{Olmos01}\\
\hline
$\chi^2/d.o.f.$ &2.87&
                 2.83&1.14 \\
\hline
$\bar{\alpha}_E\;\,[10^{-4}\,\mathrm{fm}^3]$&$11.52\pm2.43$
                 &$11.04\pm1.36$
                 &$12.4\pm0.6(\mathrm{stat})\mp0.5(\mathrm{syst})\pm0.1(\mathrm{mod})$\\
$\bar{\beta} _M\;\,[10^{-4}\,\mathrm{fm}^3]$&$ 3.42\pm1.70$
                 &$2.76\mp1.36$
                 & $1.4\pm0.7(\mathrm{stat})\pm0.4(\mathrm{syst})\pm0.1(\mathrm{mod})$\\
$b_1              $&$4.66\pm0.14$ 
                   &$4.67\pm0.14$&\\
\hline
\end{tabular}
\caption{Values for $\bar{\alpha}_E$, $\bar{\beta}_M$ and  
         $b_1$ from a fit to MAMI- and SAL-data, compared to the results from 
         \cite{Olmos01}. Note that the definition of $\chi^2/d.o.f.$ used 
         in~\cite{Olmos01}
         is different from Eq.~(\ref{eq:chisquared}). The error bars in our fits are only statistical.}
\label{table2}
\end{center}
\end{table}

\subsubsection{Static Polarizabilities}
\label{subsubsec:static}

The spin-independent static dipole polarizabilities to leading-one-loop order in SSE 
consist of the following individual contributions:
\begin{align}
  \bar{\alpha}_E&= \frac{5\,\alpha\,g_A^2} { 96\,F_\pi^2\,m_\pi\,\pi}\,
   \left(1-\frac{m_\pi}{M}\,\frac{1}{\pi}\right)
  -\frac{ 2\,\alpha\,\left(g_{117}+2\,g_{118}\right)}{ (4\,\pi\,F_\pi)^2\, M}
   \nonumber\\
&  +\frac{ \alpha\,g_{\pi N \Delta_0}^2} { 54\,\left(F_\pi\,\pi\right)^2}\,
  \left[\frac{9\,\Delta_0}{\Delta_0^2-m_\pi^2}+
  \frac{\Delta_0^2-10\,m_\pi^2}{(\Delta_0^2-m_\pi^2)^{3/2}}\,\ln R\right]
   \nonumber\\
  &= \left[11.87\,(N\pi)- (5.92\pm1.36)\,(\mathrm{c.t.})+0\,(\Delta-\mathrm{pole})
   +5.09\,(\Delta\pi)\right]\times 10^{-4}\;\mathrm{fm}^3\nonumber\\
  &=(11.04\pm1.36)\times 10^{-4}\;\mathrm{fm}^3\;,\label{alpharesult}\\
\bar{\beta}_M &= \frac{ \alpha\,g_A^2} {192\,F_\pi^2\,m_\pi\,\pi}
  +\frac{4\,\alpha \, g_{118}} { (4\,\pi\,F_\pi)^2\, M} +\frac{2\,\alpha\,
    b_{1}^2}{9\,\Delta_0 \,M^2} +\frac{ \alpha\,g_{\pi N \Delta_0}^2} {
    54\,\left(F_\pi\,\pi\right)^2}
  \frac{1}{\sqrt{\Delta_0^2-m_\pi^2}}\,\ln R\nonumber\\
  &= [ 1.25\,(N\pi)-(10.68\pm1.17)\,(\mathrm{c.t.})+(11.33\pm0.70)\,
   (\Delta-\mathrm{pole})+0.86\,(\Delta\pi)]\times 10^{-4}\;\mathrm{fm}^3\nonumber\\
  & =(2.76\mp1.36)\times 10^{-4}\;\mathrm{fm}^3\;, \label{betaresult}
\end{align}
where $ R=\left(\Delta_0+\sqrt{\Delta_0^2-m_\pi^2}\right)/{m_\pi}$ is a dimension-less parameter \cite{HHK97}. 

In the case of $\bar{\alpha}_E$, one notices a strong cancellation between the $\pi\Delta$-contributions and the short distance physics contained in $g_{117},\,g_{118}$. In Sect.~\ref{sec:comparison}, we will demonstrate that this mutual cancellation holds throughout the low energy region also in the case of the dynamical electric dipole polarizability, forcing us to the not-so surprising conclusion that for photon energies far below on-shell $\Delta\pi$ intermediate states such kind of contributions are indistinguishable from counterterms parameterizing the short distance physics. We note that the extra, quark-mass-independent term in the $\pi N$ contribution arises from our pion-threshold correction discussed in Sect.~\ref{subsec:hbchpt}.

In $\bar{\beta}_M$, we encounter the well-known cancellation between a large paramagnetism from the $\Delta$(1232) pole contributions and the nucleon-diamagnetism, which here is parameterized via the coupling $g_{118}$. In contrast to the cancellation in $\bar{\alpha}_E$ discussed above, the sum of dia- and paramagnetic effects is strongly energy-dependent and therefore leads to a clear signal in the dynamical magnetic dipole polarizability $\beta_{M1}(\omega)$, see Sect.~\ref{sec:comparison}. Apart from the contribution proportional to $g_{118}$, Eq.~(\ref{betaresult}) agrees with the result found in Ref.~\cite{HHK97} (modulo the different convention for the coupling $b_1$), where it was already noted that the $\Delta\pi$ contributions to $\bar{\beta}_M$ are sizeably smaller than in the case of $\bar{\alpha}_E$. 

From this discussion, one can already see that the two extra terms $g_{117},\,g_{118}$ are not just small higher order effects. For a consistent description both of the data and of the static polarizabilities, they are in contradistinction required in a leading-one-loop SSE analysis. Translating the fit results of Table~\ref{table2} back to these two unknown couplings, one obtains
\begin{align}
g_{117}&=17.44\pm2.11\,,\;\;g_{118}=-5.64\pm0.88\;\;\;\;(\mathrm{3-parameter-fit})\;,
\nonumber\\
g_{117}&=18.82\pm0.79\,,\;\;g_{118}=-6.05\mp0.66\;\;\;\;(\mathrm{2-parameter-fit}).
\end{align}
Therefore, these two couplings are significantly larger than their ``natural'' values, which in the Lagrangean employed here in Eqs.~(\ref{LCT1}, \ref{LCT2}) would be expected to be unity. These couplings---though formally being part of the next-to-leading one loop order Lagrangean---therefore break the naive power-counting underlying SSE and have to be taken into account already at leading-one-loop order. Having determined $g_{117},\;g_{118}$ from fits to Compton scattering data, we now have fixed all our unknown parameters and have full predictive power in the 
determination of the dynamical polarizabilities discussed in Sect.~\ref{sec:dynpols}. The addition of these two couplings does not lead to any inconsistencies with the leading-one-loop order regularization/renormalization procedure.

Finally, we note again that not only the energy dependence of the dynamical polarizabilities is independent of the two extra couplings $g_{117},\,g_{118}$, but also the values of the four spin-dependent static dipole polarizabilities $\bar{\gamma}_{E1E1}$, $\bar{\gamma}_{M1M1}$, $\bar{\gamma}_{E1M2}$, $\bar{\gamma}_{M1E2}$. The results obtained in Ref.~\cite{HHKK} are therefore reproduced\footnote{We note that in the case of $\bar{\gamma}_{E1E1}$ we obtain a small extra term $\delta\,\bar{\gamma}_{E1E1}=-\frac{\alpha g_A^2}{96 F_\pi^2\pi M m_\pi}$ due to our correction of the pion-threshold discussed in point 1 of Sect.~\ref{subsec:hbchpt}. This term is part of the next-to-leading one loop order contributions to this polarizability discussed in Ref.~\cite{CD00}.}, as expected. For better comparison with Dispersion Theory and experiment, we present here the numbers for the linear combinations $\gamma_0,\,\gamma_\pi$ of Eq.~(\ref{gpi}) in Table~\ref{table3}. For more detail, we refer the interested reader to the extensive literature on these two elusive structures \cite{Report}.

\begin{table}[!htb]
\begin{center}
\begin{tabular}{|c||c|c|c|}
\hline
Quantity&SSE&experiment&fixed-$t$ DR\\
\hline
$\gamma_0^{(p)}  \;\,[10^{-4}\,\mathrm{fm}^4]  $&$ 0.62\mp0.25$
&$-1.01\pm0.08(\mathrm{stat})\pm0.10(\mathrm{syst})$&$-0.7$\\
$\gamma_\pi^{(p)}\;\,[10^{-4}\,\mathrm{fm}^4]$&$ 8.86\pm0.25$
&$10.6\pm2.1(\mathrm{stat})\mp0.4(\mathrm{syst})\pm0.8(\mathrm{mod})$
& $\;\;\;9.3$\\
\hline
$\gamma_0^{(n)}  \;\,[10^{-4}\,\mathrm{fm}^4]$&$0.62\mp0.25$
&---&$-0.07$\\
$\gamma_\pi^{(n)}\;\,[10^{-4}\,\mathrm{fm}^4]$&$8.86\pm0.25$
&---& $\;\;\;13.7$\\
\hline
\end{tabular}
\caption{Comparison of proton $\gamma_0^{(p)}$, $\gamma_\pi^{(p)}$ and neutron $\gamma_0^{(n)}$, $\gamma_\pi^{(n)}$ spin-polarizabilities between leading-one-loop SSE, experiment and Hyperbolic Dispersion Theory \cite{Report}; the experimental values for $\gamma_0^{(p)}$, $\gamma_\pi^{(p)}$ are taken from \cite{GDH} and \cite{Olmos01}, respectively. Note that the $\mathcal{O}(\epsilon^3)$ SSE-results are purely isoscalar.}\label{table3}
\end{center}
\end{table}
%\begin{table}[!htb]
%\begin{center}
%\begin{tabular}{|c||c|c|}
%\hline
%Quantity&SSE&DR\\
%\hline
%$\gamma_0^{(s)}  $&$ (0.62\mp0.25)\cdot10^{-4}\,\mathrm{fm}^4$
%&$-0.42\cdot10^{-4}\,\mathrm{fm}^4$\\
%$\gamma_\pi^{(s)}$&$ (8.86\pm0.25)\cdot10^{-4}\,\mathrm{fm}^4$
%&$11.95\cdot10^{-4}\,\mathrm{fm}^4$\\
%\hline
%$\gamma_0^{(v)}  $&$0+{\cal O}(\epsilon^4)$&$-0.33\cdot10^{-4}\,\mathrm{fm}^4$\\
%$\gamma_\pi^{(v)}$&$0+{\cal O}(\epsilon^4)$&$-1.65\cdot10^{-4}\,\mathrm{fm}^4$\\
%\hline
%\end{tabular}
%\caption{Comparison of isoscalar $\gamma_0^{(s)}$, $\gamma_\pi^{(s)}$ and isovector 
%$\gamma_0^{(v)}$, $\gamma_\pi^{(v)}$ spin-polarizabilities between leading-one-loop 
%SSE and Dispersion Theory}\label{table3}
%\end{center}
%\end{table}

Our findings on the static quadrupole polarizabilities $\bar{\alpha}_{E2},\,\bar{\beta}_{M2}$ are discussed in App.~\ref{sec:quadrupole}. We now move on to a detailed discussion of the dynamical polarizabilities.

%%%%%%%%%%%%%%% Intro %%%%%%%%%%%%%%%%%%%
\section{Chiral Dynamics and Dynamical Polarizabilities}
\setcounter{equation}{0}
\label{sec:comparison}
%%%%%%%%%%%%%%%%%%%%%%%%%%%%

%%%%%%%%%%%%%%%%%%%%%%%%%%%%
\subsection{Isoscalar Spin-Independent Polarizabilities}
\label{sec:dynpols}

\begin{figure}[!htb]
  \begin{center}
    \includegraphics*[width=0.99\textwidth]{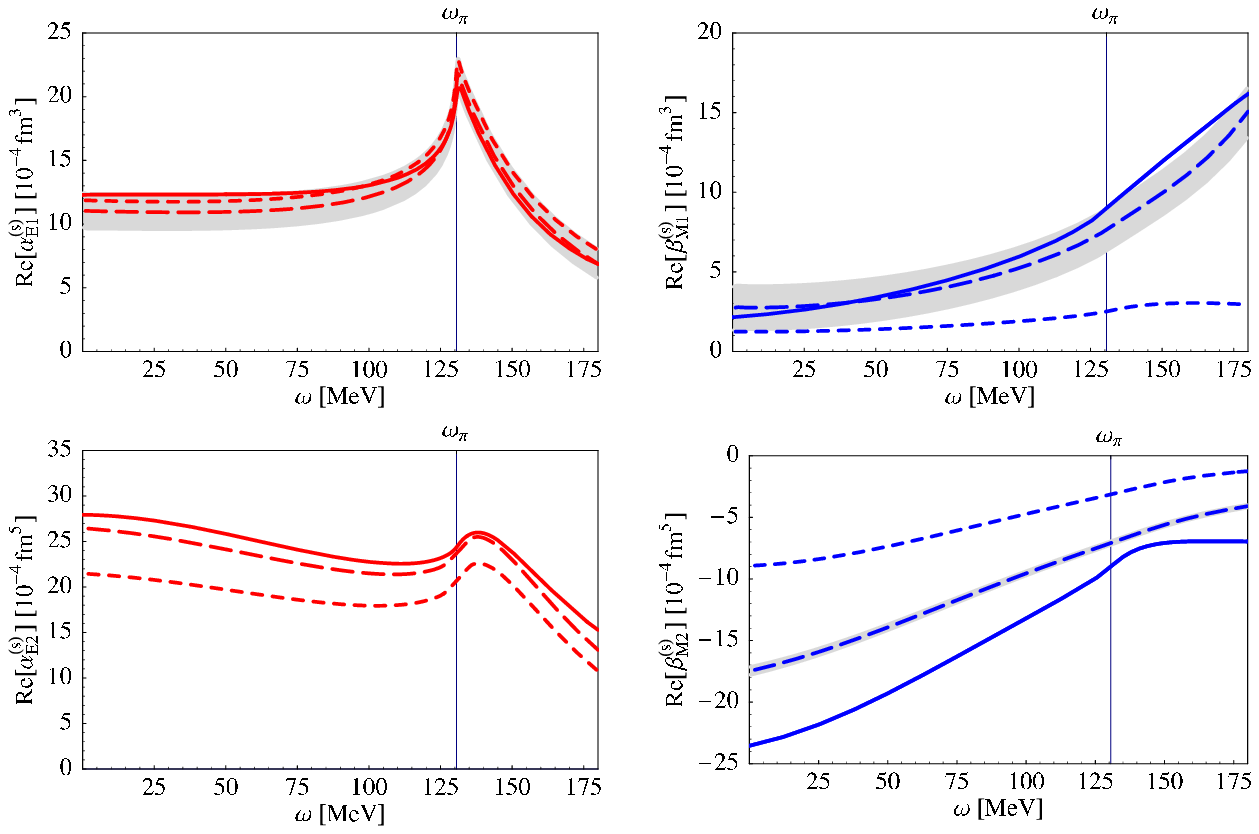}
 \caption{Comparison of the Dispersion Theory results (solid line) of the
   real parts of the isoscalar, spin-independent dynamical electric and
   magnetic dipole (top) and quadrupole (bottom) polarizabilities with
   leading-one-loop order HB$\chi$PT (short dashed line) and SSE (long dashed line) respectively.
   The grey band around the SSE curves arises from the uncertainty in the fit parameters 
   determined with the help of the Baldin sum rule in Sect.~\ref{subsec:fit}.}\label{fig3}
  \end{center}
\end{figure}

Turning first to $\alpha_{E1}^{(s)}(\omega)$ as shown in
Fig.~\ref{fig3}, it is obvious that its energy dependence in the low energy region is entirely controlled by chiral dynamics arising from single $\pi N$ intermediate states. All three theoretical analyses agree rather well within the statistical uncertainty band of the SSE calculation. As already discussed for the static electric polarizability $\bar{\alpha}_E$ in the previous section, no effects from any inherent $\pi\Delta$ intermediate states can be detected, pointing to the fact that these rather heavy degrees of freedom are effectively frozen out at these low energies. This makes them---as far as the energy dependence of the dynamical polarizabilities is concerned---indistinguishable from short distance contributions contained in the couplings $g_{117},\, g_{118}$. We also note that the strength and shape of the cusp associated with the one-pion production threshold is reproduced extremely well by the leading-one-loop chiral calculations. It will serve as an interesting check for the
convergence properties of the chiral theories to see whether the rather good agreement is maintained, once the higher order corrections are included. 

The other spin-independent $l=1$ dynamical polarizability, $\beta_{M1}^{(s)}(\omega)$, shows quite a different picture. We note that the three theoretical frameworks only agree (within the uncertainty of the SSE parameters) for the value of the static magnetic polarizability $\bar{\beta}_M$. For increasing values of the photon energy, it becomes obvious from the agreement between SSE and Dispersion Theory that explicit $\Delta$(1232) contributions via $s$-channel pole graphs lead to a paramagnetic behaviour quickly rising with energy. Any $\Delta\pi$ contributions remain small and are effectively frozen out.  The near cancellation between para- and diamagnetic contributions for the static value discussed in the previous section is completely taken over by $\Delta$(1232)-induced paramagnetism when the photon energy goes up. We explicitly point to the scale on the $y$-axis of this plot, indicating a rise by a factor of four at photon energies near the one-pion production threshold. While the leading-one-loop HB$\chi$PT calculation \cite{BKKM92} provides a good prediction for $\bar{\beta}_M$, it fails to describe the energy dependence of $\beta_{M1}^{(s)}(\omega)$, as shown in Fig.~\ref{fig3}. In contrast to $\alpha_{E1}^{(s)}(\omega)$, hardly any cusp is visible in $\beta_{M1}^{(s)}(\omega)$. Beyond the static limit, the chiral $\pi N$ contributions play a minor role in this polarizability. We note that while the fine details of the rising paramagnetism in $\beta_{M1}^{(s)}(\omega)$ differ between SSE and Dispersion Theory, they are consistent within the uncertainties of the SSE curve. The discrepancy between the two schemes above the one-pion production threshold is likely to be connected to a detailed treatment of the width of the $\Delta$-resonance, which is neglected in leading-one-loop SSE. 

We further note that the good agreement between SSE and Dispersion Theory for the $l=1$ spin-independent dynamical polarizabilities provides a non-trivial check regarding the physics parameterized in the couplings $g_{117},\,g_{118}$. Given that these two structures are energy independent, cf. 
Eqs.~(\ref{LCT1}, \ref{LCT2}), the fact that only the $\pi N$ and $\Delta$ degrees of freedom suffice to describe the energy dependence in the low energy region quite well supports our idea that the physics underlying $g_{117},\, g_{118}$ is ``short-distance'' from the point of view of $\chi$EFTs. 

It is also interesting to look at the spin-independent $l=2$ dynamical polarizabilities, even if in actual analyses of Compton data they only play a minor role. In $\alpha_{E2}^{(s)}(\omega)$ we observe a visible contribution from $\Delta\pi$ intermediate states. It hardly modifies the shape of the energy dependence, but does affect the overall normalization of this polarizability, as can be seen from the difference between the SSE and the HB$\chi$PT curve. The agreement between SSE and Dispersion Theory is surprisingly good throughout the entire low energy region. Another interesting higher order dynamical polarizability is $\beta_{M2}^{(s)}(\omega)$. The chiral $\pi N$ contribution seems to play only a minor role in the energy dependence of this polarizability. $\Delta\pi$ and a surprisingly large $\Delta$(1232) $u$-channel pole contribution can close a significant part of the gap between the HB$\chi$PT and the Dispersion Theory result. The remaining gap between SSE and Dispersion Theory might well be accounted for by next-to-leading one loop chiral $\pi N$ corrections, given that the slope of the energy dependence below pion threshold seems to agree between the two frameworks. Nevertheless, the energy dependence of this polarizability is quite peculiar. The magnetic quadrupole strength has decreased rather fast by more than a factor of two when the photon energy reaches the one-pion production threshold. This shape is reminiscent of a relaxation effect typically discussed in textbook examples for dispersive effects \cite{GH1}. While both in HB$\chi$PT and SSE the strengths for 
$\beta_{M2}(\omega)$ tend to zero for large photon energies, the DR-curve seems to point to additional physics contributions above the pion threshold.

We now move on to a discussion of the $l=1$ spin-dependent dynamical polarizabilities.

%%%%%%%%%%%%%%%%%%%%%%%%%%%%
\subsection{Isoscalar Spin-Dependent Polarizabilities}
%%%%%%%%%%%%%%%%%%%%%%%%%%%%%%%%
  
\begin{figure}[!htb]
  \begin{center}
    \includegraphics*[width=0.99\textwidth]{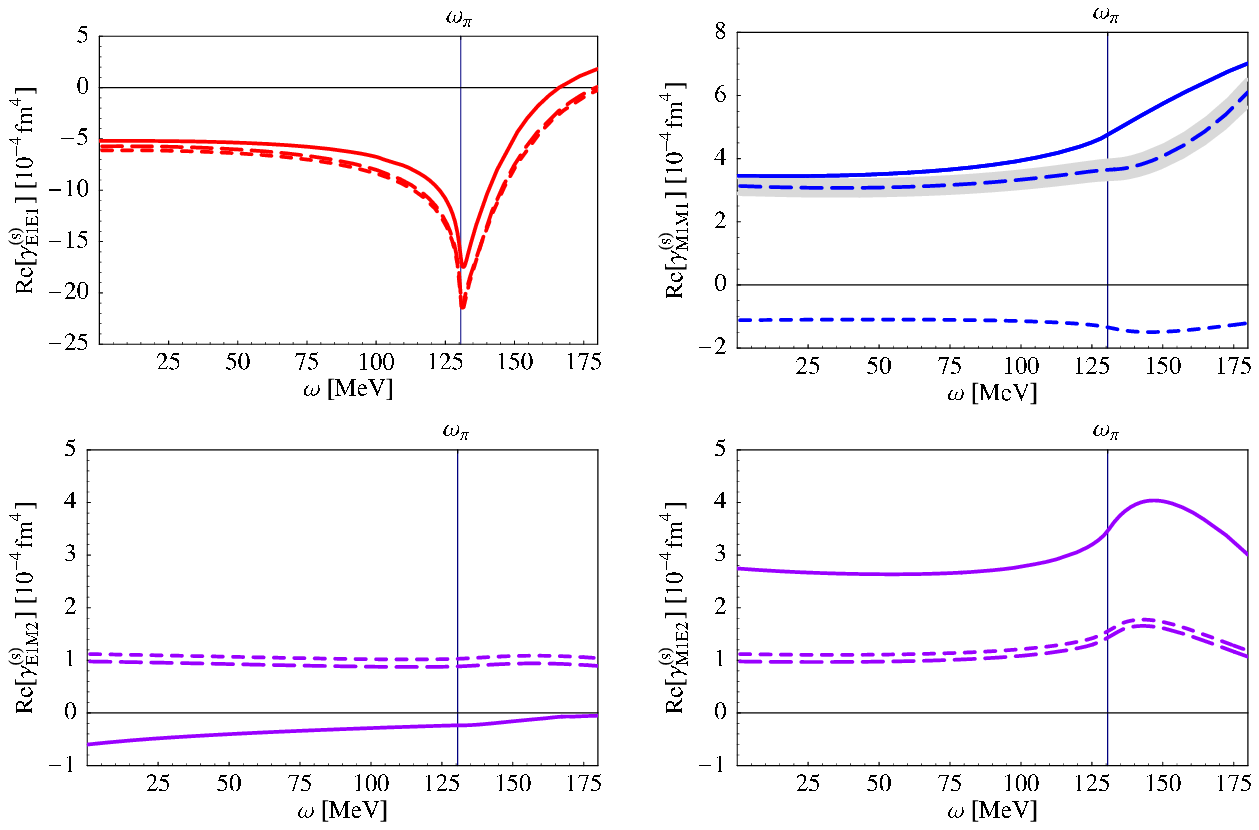}
 \caption{Comparison of the Dispersion Theory results of the
   real parts of the isoscalar, spin-dependent dynamical dipole
   polarizabilities with leading-one-loop HB$\chi$PT, respectively SSE. Notation as in 
Fig.~\ref{fig3}.}\label{fig4}
  \end{center}
\end{figure}

We again remind the reader that no fit parameters analogous to $g_{117}$ and
$g_{118}$ are present in the leading-one-loop SSE results for the spin-dependent 
polarizabilities. The only free parameter entering the dynamical spin polarizabilities is $b_1$, which we have determined from the fit to 
Compton cross sections in Sect.~\ref{subsec:fit}. 
As Fig.~\ref{fig4} demonstrates, the contributions of the $\Delta\pi$
continuum to the spin polarizabilities are small throughout the low energy region. The energy dependence in $\gamma_{E1E1}^{(s)}(\omega)$ is completely governed by chiral dynamics and agrees well among the three frameworks, quite analogous to the situation in $\alpha_{E1}^{(s)}(\omega)$. The $\Delta$(1232) pole contribution---rising with energy---is visible in $\gamma_{M1M1}^{(s)}(\omega)$, but it does not rise as dramatically as in the case of $\beta_{M1}^{(s)}(\omega)$ (cf. Fig.~\ref{fig3}). The HB$\chi$PT calculation for $\gamma_{M1M1}^{(s)}(\omega)$ deviates strongly from both the SSE and DR result, signaling again the need for explicit $\Delta$(1232) degrees of freedom in resonant multipoles. The slight disagreement between SSE and DR for photon energies above pion threshold in $\gamma_{M1M1}^{(s)}(\omega)$ might be connected to a detailed treatment of the width of the resonance, see Sect.~\ref{sec:future}. Both HB$\chi$PT and SSE predictions for the mixed spin polarizabilities are rather similar, disagreeing with the DR result. While $\gamma_{E1M2}^{(s)}(\omega)$ constitutes a rather tiny structure effect which will be hard to pin down precisely, the ``large'' gap between SSE and the DR result in $\gamma_{M1E2}^{(s)}(\omega)$ could possibly arise from the missing $E2$ excitation of the $\Delta$ resonance in a leading-one-loop SSE calculation. This effect can be accounted for at next-to-leading one loop order. On the other hand, the overall shape of the energy dependence in $\gamma_{M1E2}^{(s)}(\omega)$ is rather similar between the chiral and the DR results, indicating that a $\pi N$ loop contribution at the next higher chiral order might also suffice to close the gap. 

In conclusion, among the four isoscalar \hspace{.01cm} spin-dependent dipole polarizabilities, only  $\gamma_{E1E1}^{(s)}(\omega)$ seems to be dominated by $\pi N$ chiral dynamics, which can be accounted for rather well already at leading-one-loop order throughout the low energy region. A detailed understanding of the dynamical dipole spin polarizabilities requires explicit $\Delta$(1232) resonance degrees of freedom. 

After this discussion focused on the low-energy tail of Compton-scattering, we turn to 
a short excursion on dynamical polarizabilities in the resonance region in the next section.

%%%%%%%%%%%%%%%%%%%%%%%%%%%
\section{Dynamical Polarizabilities in the Resonance Region}
\setcounter{equation}{0}
\label{sec:future}
%%%%%%%%%%%%%%%%%%%%%%%%%%%%%%%

\begin{figure}[!htb]
  \begin{center}
    \includegraphics*[width=0.3\textwidth]{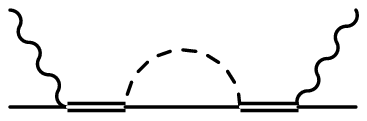}
 \caption{Schematic representation of the intermediate $\pi N$ states generating the width of the $\Delta$ resonance.}\label{Deltawidth}
  \end{center}
\end{figure}

\begin{figure}[!htb]
  \begin{center}
    \includegraphics*[width=0.99\textwidth]{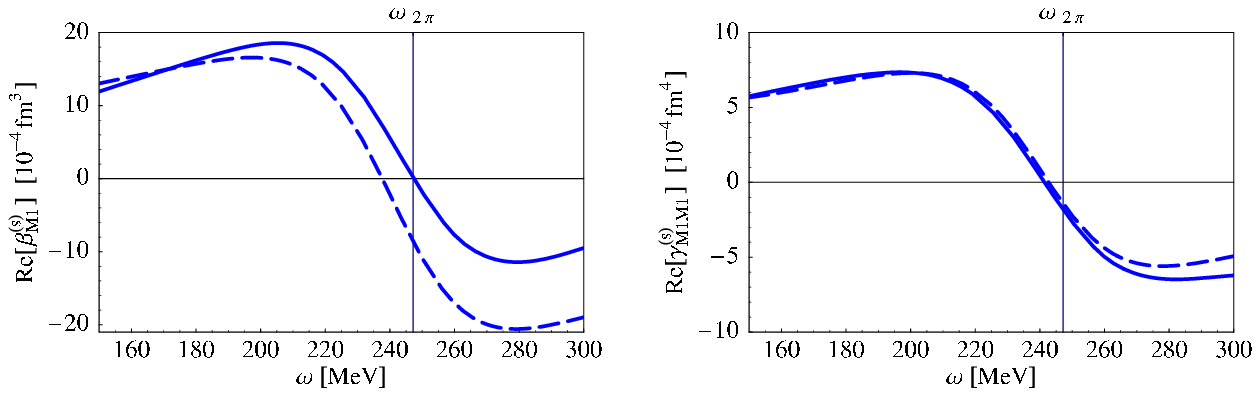}
 \caption{Qualitative comparison between Dispersion Theory (solid lines) and a $\chi$EFT calculation employing spin 3/2 propagators with an additional Breit-Wigner parameterization (dashed lines) for two selected dynamical polarizabilities in the resonance region.}\label{SSEresonant}
  \end{center}
\end{figure}

We remind the readers that in SSE, the finite width $\Gamma_\Delta/2=50$ MeV of the $\Delta$(1232) resonance \cite{HDT} is treated as a (small) perturbation. Clearly, as soon as the expression $(W-M-\Delta_0)$ in the denominator of the $\Delta(1232)$ pole contributions in Eqs.~(\ref{A1})--(\ref{A6}) becomes comparable in magnitude to $\Gamma_\Delta$, this assumption breaks down. Therefore, the SSE results become meaningless for photon energies above 180 MeV and we have truncated the plots in Sects.~\ref{subsec:experiment} and \ref{sec:comparison} at this energy. When one wants to analyze dynamical polarizabilities within $\chi$EFT at such higher photon energies---i.e. in the first or second resonance region---one clearly has to modify the power-counting employed by resumming the $\pi N$-intermediate state contribution (Fig.~\ref{Deltawidth}) to the $\Delta$ self energy to all orders. While a full calculation of nucleon Compton scattering in the resonance region in appropriately modified SSE is quite involved, one can get a quick qualitative picture of the results in the resonant multipole channels by adding (by hand) a Breit-Wigner parameterization to the spin 3/2 propagators. In Fig.~\ref{SSEresonant} we show the results of such a procedure for the (isoscalar) magnetic polarizabilities $\beta_{M1}(\omega)$ and $\gamma_{M1M1}(\omega)$. While details of the comparison to the Dispersion Theory result are certainly far from perfect, one realizes that such a procedure does lead to a good qualitative description ({\it e.g.} observe the zero crossing of the energy dependence on the right shoulder of the $\Delta$ resonance) of the dynamical polarizabilities even in the resonance region---without of course replacing the need for a full systematic calculation in $\chi$EFT \cite{future}.

\begin{figure}[!htb]
  \begin{center}
    \includegraphics*[width=0.9\textwidth]{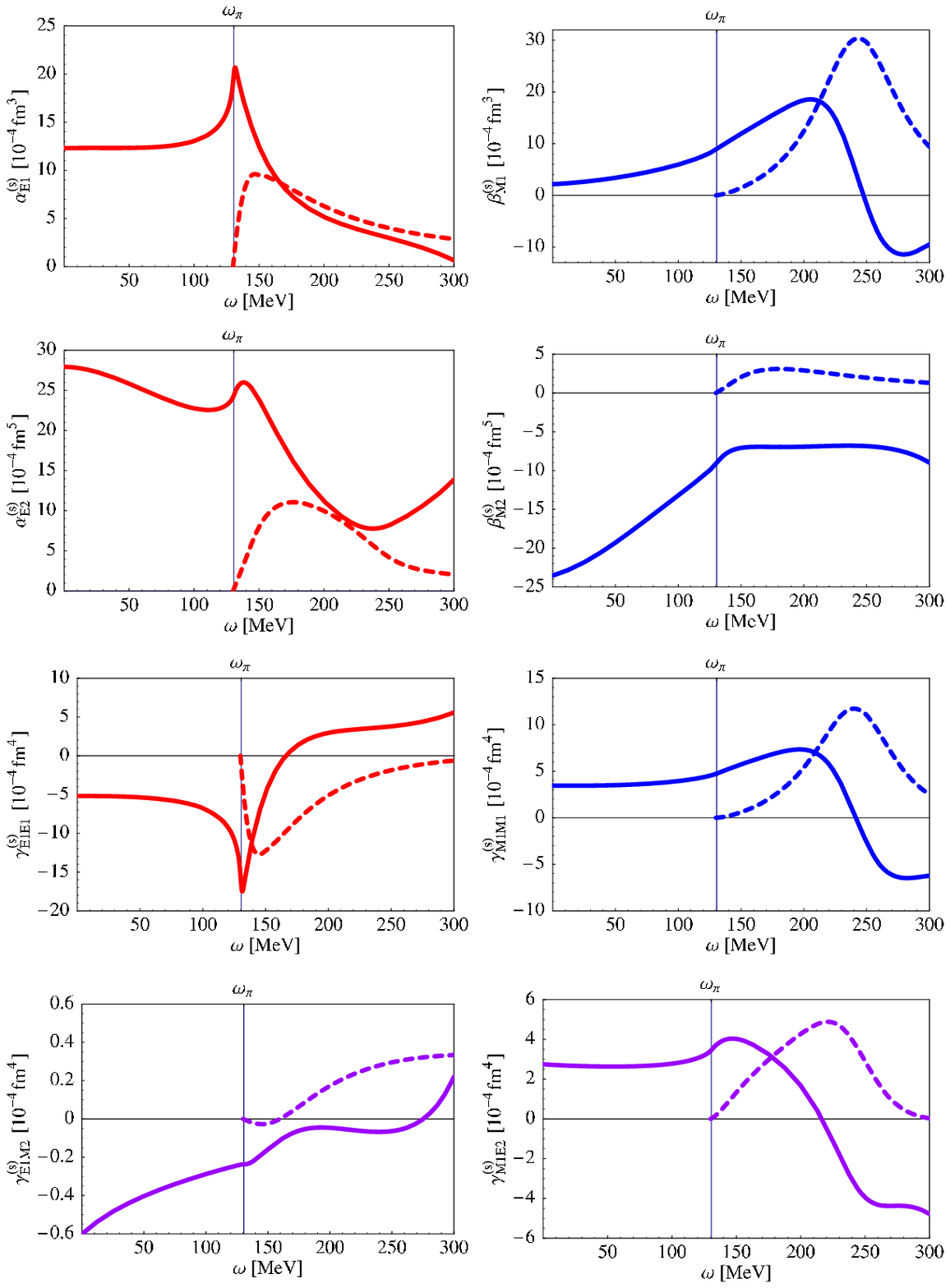}
 \caption{Dispersion Theory result 
   for the real (solid lines) and imaginary parts (dashed lines) of the isoscalar dynamical polarizabilities discussed in the text.}\label{fig1a}
  \end{center}
\end{figure}

Obviously, systematic predictions for dynamical polarizabilities from Dispersion Theory are not limited to photon energies below the $\Delta$-resonance. In Fig.~\ref{fig1a}, we show the DR prediction for the first 8 isoscalar dynamical polarizabilities throughout the first and second resonance region. For comparison we also plot the (surprisingly) large contributions from the imaginary parts of the dynamical polarizabilities in this region, which below the two-pion threshold -- corresponding to $\omega=247.2$ MeV in the cm system -- in DR are input obtained from Ref.~\cite{HDT}. 
Above the two-pion threshold, they are modeled with the same input which enters the imaginary parts of the amplitudes $A_i^L$ in the $s$-channel (see Sect. \ref{sec:dispersionformalism}) 
%While the DR predictions seem to be well behaved in the entire energy region, we point out that above the two-pion threshold---corresponding to $\omega=247.2$ MeV in the cm system---the model dependence implicit in the DR results increases considerably. 

\begin{figure}[!htb]
  \begin{center}
    \includegraphics*[width=0.9\textwidth]{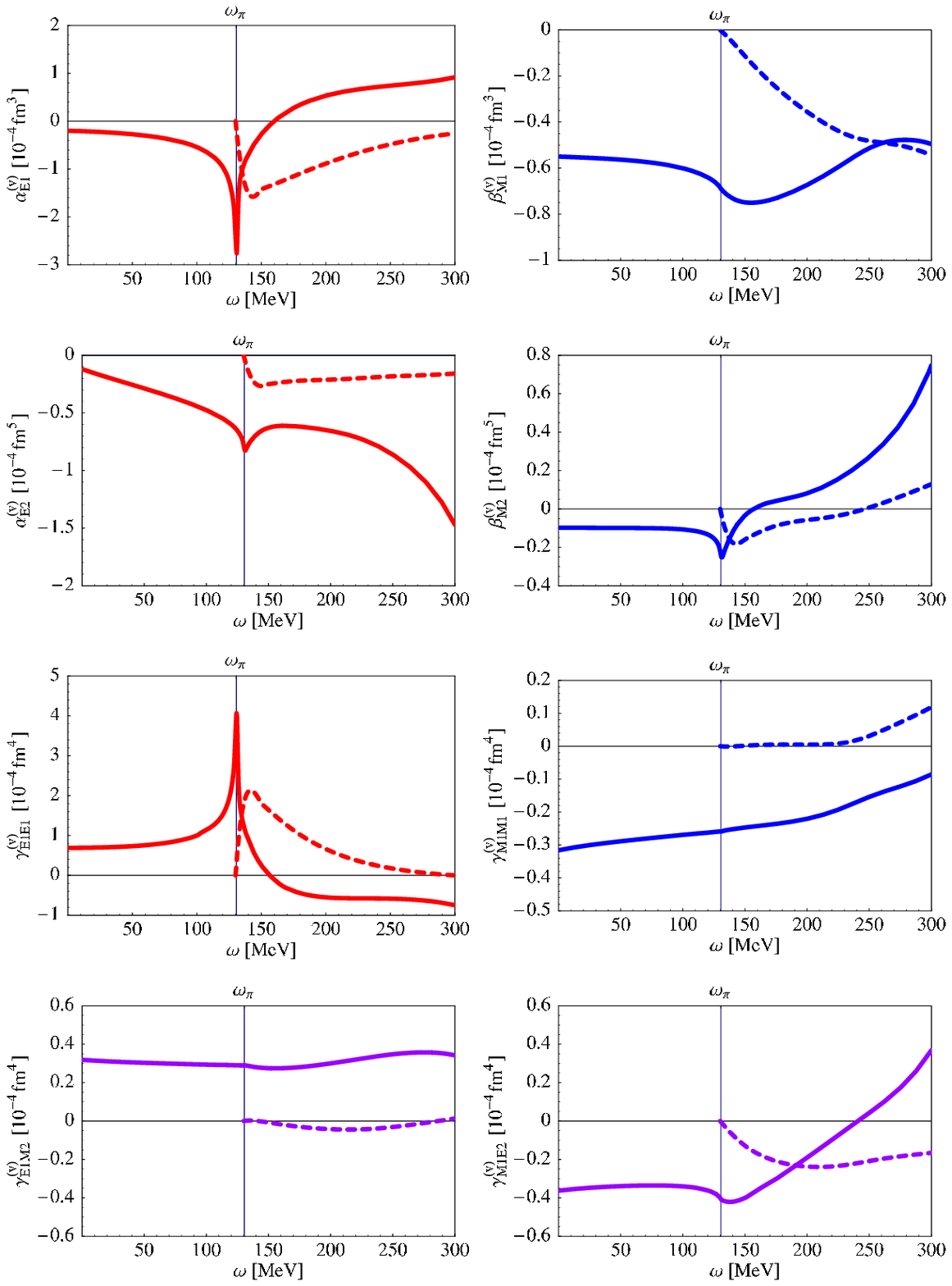}
 \caption{Dispersion Theory result 
   for the isovector dynamical polarizabilities. For notation see Fig.~\ref{fig1a}.}\label{fig2a}  
  \end{center}
\end{figure}

Likewise, we show in Fig.~\ref{fig2a} the predictions from Dispersion Theory for the first 8 isovector dynamical polarizabilities of the nucleon from low energies up into the second resonance region. We note that the isovector contributions---defined as half the difference between proton and neutron results---are a lot smaller than their isoscalar counterparts. This finding is in agreement with $\chi$EFT, which finds a null result for all isovector polarizabilities to leading-one-loop order \cite{HHK97,HHKK}.  The isovector-dependent terms are treated as small higher order corrections, both in HB$\chi$PT and in SSE. 
Before concluding we note that there seems to be an exception to this rule---according to our results from Dispersion Theory: The spin-dependent dynamical polarizability $\gamma_{E1M2}^{(v)}(\omega)$ is found to be of the same size as its isoscalar counterpart $\gamma_{E1M2}^{(s)}(\omega)$ in Fig.~\ref{fig1a}, which however is significantly smaller than the other iso-scalar dipole polarizabilities.

%%%%%%%%%%%%%%% Conclusion %%%%%%%%%%%%%%%%%%%
\section{Conclusions}
\setcounter{equation}{0}
\label{sec:con}
%%%%%%%%%%%%%%%%%%%%%%%%%%%%

Dynamical polarizabilities are a
concept complementary to \emph{generalized} polarizabilities of the nucleon \cite{Guichon, genpoleff, genpolDR}. The latter probe
the nucleon in virtual Compton scattering, i.e. with an incoming photon of
non-zero virtuality  and an outgoing, static real
photon. Therefore, they provide information about the spatial distribution of charges and magnetism  inside the nucleon at zero energy. Dynamical polarizabilities---the main subject of this work---on the other hand test the
global low energy excitation spectrum of the nucleon at {\em non-zero} energy and answer the question, which internal degrees of freedom govern the structure of the nucleon at low energies.

In this work, we have confronted the concept of dynamical polarizabilities developed in Ref.~\cite{GH1} with data from nucleon Compton scattering experiments via a multipole expansion. We compared $\chi$EFT and Dispersion Theory predictions for the dynamical polarizabilities. The pertinent results of our analysis can be summarized as follows:
\begin{itemize}
\item [i)] Both state-of-the-art Dispersion Theory as well as chiral effective field theory with explicit $\Delta$(1232) degrees of freedom are able to describe the proton Compton cross sections in the low energy region very well.  
We note that without explicit $\Delta(1232)$ degrees of freedom, the calculations deviate considerably at backward angles for photon energies starting as early as 80 MeV. 
\item[ii)] An $l=1$ truncation in the multipole expansion of Compton scattering is found to be completely sufficient to describe available cross sections up to energies of the $\Delta$ resonance. This implies that all information about the structure of the nucleon in these low energy processes is contained in the six dynamical dipole polarizabilities.
\item[iii)] There is an impressive agreement between Dispersion Theory and SSE in most multipole channels. Differing predictions in some dynamical polarizabilities cannot be resolved at the present level of experimental error bars in proton Compton scattering.
\item[iv)] We have identified the dynamical dipole polarizabilities $\alpha_{E1}(\omega)$ and $\gamma_{E1E1}(\omega)$ as candidates whose entire energy dependence in the low energy domain is controlled by chiral dynamics. 
\item[v)] A projector formalism both for Dispersion Theory and for $\chi$EFT allows one to uniquely generate any desired dynamical polarizability from a given set of structure-dependent Compton amplitudes.
\item[vi)] We have determined the static electric and magnetic polarizabilities of the proton within SSE via a fit to Compton cross section data at all angles up to $\omega=170\;\mathrm{MeV}$. The results are entirely consistent with results one obtains when Dispersion Theory techniques are employed to extract static polarizabilities of the nucleon from Compton data. Our SSE-amplitudes (App.~\ref{sec:AppB}) therefore provide an alternative extraction method for these static polarizabilities directly from Compton data up to 170~MeV.
\end{itemize}
%We conclude our present work on dynamical polarizabilities in nucleon Compton scattering with the remark, that due to item ii), an ideal high precision measurement of spin-averaged Compton scattering at fixed photon energy and more than six angles could dis-entangle the contributions from the various multipoles. Doing that, one could then directly insert data points into the plots for the dynamical dipole polarizabilities (Figs.~\ref{fig3}, \ref{fig4}). 
In an upcoming analysis~\cite{polarized}, we will investigate how to determine all $l=1$ dynamical polarizabilities at a given energy directly from nucleon Compton scattering experiments without resorting to any theoretical machinery like SSE or Dispersion Theory.

%%%%%%%%%%%%%%%%%%%%%%%%%%%%%%%%%%%%%%%%%%%%%%%%%%%%%%%%%%%%%%%%%%%%%%%%%%%%%%%
%%%%%%%%%%%%%%%%%%%%%%%%%%%%%%%%%%%%%%%%%%%%%%%%%%%%%%%%%%%%%%%%%%%%%%%%%%%%%%%

\section*{Acknowledgments}

The authors acknowledge helpful discussions with N. Kaiser, M. Schumacher and W. Weise. HWG, TRH, BP and RPH are grateful to the ECT* in Trento for its hospitality where a large part of the work was done. This work has been supported in part by the Bundesministerium f\"ur Forschung und Technologie, and by Deutsche Forschungsgemeinschaft under contract GR1887/2-1 (HWG and RPH).

\newpage

%%%%%%%%%%%%%%%%%%%%%%%%%%%%%%%%%%%%%%%%%%%%%%%%%%%%%%%%%%%%%%%%%%%%
\appendix
%%%%%%%%%%%%%%%%%%%%%%%%%%%%%%%%%%%%%%%%%%%%%%%%%%%%%%%%%%%%%%%%%%%%

%%%%%%%%%%%%%%% Intro %%%%%%%%%%%%%%%%%%%
\section{Projection Formulae in Dispersion Theory}
\setcounter{equation}{0}
\label{sec:AppA}
%%%%%%%%%%%%%%%%%%%%%%%%%%%%
In this appendix, we give the relevant formulae to calculate the multipole
amplitudes for Compton scattering from the invariant amplitudes $A_i^L$.
Following the notation of Ref.~\cite{Pfeil}, we introduce the following six
independent helicity amplitudes $\phi_{\Lambda'\Lambda}$, with
$\Lambda=\lambda_\gamma-\lambda_N$ $(\Lambda'=\lambda'_\gamma-\lambda'_N$)
related to the helicities of the initial (final) photon and nucleon,
$\lambda_\gamma$ ($\lambda'_\gamma$) and $\lambda_N$ ( $\lambda'_N$),
respectively, \begin{eqnarray} \phi_1 &\equiv& \phi_{1/2\;1/2}
  \nonumber\\
  \phi_2 &\equiv& \phi_{1/2\;-1/2}
  \nonumber\\
  \phi_3 &\equiv& \phi_{1/2\;-3/2}
  \nonumber\\
  \phi_4 &\equiv& \phi_{1/2\;3/2}
  \nonumber\\
  \phi_5 &\equiv& \phi_{3/2\;3/2}
  \nonumber\\
  \phi_6 &\equiv& \phi_{3/2\;-3/2}.  \end{eqnarray}

The invariant amplitudes $A_i^L$ are connected to the helicity amplitudes
$\phi_i$ by the relations
\begin{eqnarray} & 
  \phi_1=&\frac{\sqrt{(1-\sigma)}}{8\pi\sqrt{s}}
  \frac{(s-M^2)[2(s-M^2)+t]}{2M^3[M^2\sigma-s(\sigma-2)]}\nonumber\\
  & &\times \left\{(\sigma -1)s[2M^2\; A_3^L-(s-M^2)A_4^L] +2M^2 A_6^L(\sigma
    M^2-s)\right\},
  \nonumber\\
   &\phi_2=&-\frac{\sqrt{\sigma}}{8\pi\sqrt{s}}\frac{(s-M^2)^2}{4M^2s^{3/2}}\nonumber\\
  & &\times \{-2M^2\sigma[A_1^L(s+M^2)+A_2^L(s-M^2)] +s
  A_5^L(\sigma-2)[2(s-M^2)+t]\},
  \nonumber\\
  & \phi_3=&-\frac{\sigma\, \sqrt{1-\sigma}}{8\pi\sqrt{s}}
  \frac{(s-M^2)^2}{4Ms}
  \{4M^2 A_1^L-A_5^L[2(s-M^2)+t]\},\nonumber\\
  & \phi_4=&\frac{\sqrt{\sigma}\, (1-\sigma)}{8\pi\sqrt{s}}
  \frac{\sqrt{s}(s-M^2)[2(s-M^2)+t]}{2M^2[M^2\sigma-s(\sigma-2)]}
  [2M^2 A_6^L+A_3^L(s+M^2)],\nonumber\\
  & \phi_5=&-\frac{(1-\sigma)\sqrt{(1-\sigma)}}{8\pi\sqrt{s}}
  \frac{s(s-M^2)[2(s-M^2)+t]}{M[M^2\sigma-s(\sigma-2)]}
  [A_3^L+A_6^L+A_4^L\frac{(s-M^2)}{2M^2}],\nonumber\\
  & \phi_6=&\frac{\sigma\sqrt{\sigma}}{8\pi\sqrt{s}}
  \frac{(s-M^2)^2}{4s\sqrt{s}}\{2(s-M^2) A_2^L-2 A_1^L(s+M^2) +A_5^L[2(s-M^2)
  +t]\},
\label{eq:hel_inv}
\end{eqnarray}
where $\sigma=-s\;t/(s-M^2)^2=\sin^2(\theta/2).$ 

The helicity amplitudes have the following standard partial-wave decomposition
in terms of the reduced matrices $d^J_{\Lambda\Lambda'}$
\begin{eqnarray}
\phi_{\Lambda' \Lambda}=\sum_J(2J+1)\phi_{\Lambda' \Lambda}^J 
 \; d^J_{\Lambda' \Lambda}(\theta),
\label{eq:partial_wave}
\end{eqnarray}
which, by inversion, gives
\begin{eqnarray}
\phi_{\Lambda' \Lambda}^J =\frac{1}{2}\int_{-1}^{+1}
d\cos\theta \; \; \phi_{\Lambda'
  \Lambda}(\cos\theta)  
d^J_{\Lambda' \Lambda}(\theta).
\label{eq:inv_partial_wave}
\end{eqnarray}

With the partial wave decomposition of Eq.~(\ref{eq:partial_wave}), we finally
obtain the relations between the multipole amplitudes of Compton scattering
and the helicity partial waves:
\begin{eqnarray}
f^{l+}_{E E}
&=&\frac{1}{(l+1)^2}\left[\frac{1}{2}(\phi_1^{l+1/2}-\phi_2^{l+1/2})
+\sqrt{\frac{l+2}{l}}(\phi_3^{l+1/2}-\phi_4^{l+1/2})
+      \frac{l+2}{2l}(\phi_5^{l+1/2}-\phi_6^{l+1/2})\right],
\nonumber\\
f^{l+}_{M M}
&=&\frac{1}{(l+1)^2}\left[\frac{1}{2}(\phi_1^{l+1/2}+\phi_2^{l+1/2})
-\sqrt{\frac{l+2}{l}}(\phi_3^{l+1/2}+\phi_4^{l+1/2})
+\frac{l+2}{2l}(\phi_5^{l+1/2}+\phi_6^{l+1/2})\right],
\nonumber\\
f^{l-}_{E E}
&=&\frac{1}{l^2}\left[
\frac{1}{2}(\phi_1^{l-1/2}+\phi_2^{l-1/2})
+\sqrt{\frac{l-1}{l+1}}(\phi_3^{l-1/2}+\phi_4^{l-1/2})
+\frac{l-1}{2(l+1)}(\phi_5^{l-1/2}+\phi_6^{l-1/2})\right],
\nonumber\\
f^{l-}_{M M}
&=&\frac{1}{l^2}\left[
\frac{1}{2}(\phi_1^{l-1/2}-\phi_2^{l-1/2})
-\sqrt{\frac{l-1}{l+1}}(\phi_3^{l-1/2}-\phi_4^{l-1/2})
+\frac{l-1}{2(l+1)}(\phi_5^{l-1/2}-\phi_6^{l-1/2})\right],
\nonumber\\
f^{l+}_{E M}
&=&\frac{1}{(l+1)^2}\left[-\frac{1}{2}(\phi_1^{l+1/2}-\phi_2^{l+1/2})
-\frac{1}{\sqrt{l(l+2)}}(\phi_3^{l+1/2}-\phi_4^{l+1/2})
+\frac{1}{2}(\phi_5^{l+1/2}-\phi_6^{l+1/2})\right],\nonumber\\
f^{l+}_{M E}
&=&\frac{1}{(l+1)^2}\left[-\frac{1}{2}(\phi_1^{l+1/2}+\phi_2^{l+1/2})
+\frac{1}{\sqrt{l(l+2)}}(\phi_3^{l+1/2}+\phi_4^{l+1/2})
+\frac{1}{2}(\phi_5^{l+1/2}+\phi_6^{l+1/2})\right].\nonumber\\
& &
\label{eq:phi_f}
\end{eqnarray}

%%%%%%%%%%%%%%% Intro %%%%%%%%%%%%%%%%%%%
\section{Compton amplitudes to leading-one-loop order in $\chi$EFT}
\setcounter{equation}{0}
\label{sec:AppB}
%%%%%%%%%%%%%%%%%%%%%%%%%%%%

The formulae which connect the amplitudes $R_i$ discussed in the text to the $A_i^H$ basis usually used in 
$\chi$EFT calculations of nucleon Compton scattering read \cite{babusci}
\begin{align}
  A_1^H&= 4\pi \frac{W}{M}(R_1+z R_2)  \nonumber\\
  A_2^H&=-4\pi \frac{W}{M} R_2 \nonumber\\
  A_3^H&= 4\pi \frac{W}{M}(R_3+z R_4+2z R_5+2R_6)\nonumber\\
  A_4^H&= 4\pi \frac{W}{M} R_4 \nonumber\\
  A_5^H&=-4\pi \frac{W}{M}(R_4+  R_5)  \nonumber\\
  A_6^H&=-4\pi \frac{W}{M} R_6
\label{AR}
\end{align}
As discussed in Sect.~\ref{sec:multipoles} we need to know both the pole as well as the structure-dependent contributions to $A_i^H$.

The cm pole contributions to the Compton amplitudes $A_1^H-A_6^H$ for the case of a proton target have been calculated up to leading-one-loop order in Ref.~\cite{HHKK}. For completeness we list them here again ($\kappa=\frac{1}{2}\,(\kappa_v+\kappa_s)$):
\begin{align}
  A_1^{\mathrm{pole}}(\omega,\,z)&=-\frac{e^2}{M}
  +\mathcal{O}(\epsilon^4)  \nonumber\\
  A_2^{\mathrm{pole}}(\omega,\,z)&= \frac{e^2\,\omega}{M^2}
  +\mathcal{O}(\epsilon^4)  \nonumber\\
  A_3^{\mathrm{pole}}(\omega,\,z)&=
  \frac{e^2\,\omega\,\left(1+2\,\kappa-(1+\kappa)^2\,z\right)}{2\,M^2}
  -\frac{e^2\,g_A}{4\,\pi^2\,f_\pi^2}\,\frac{\omega^3\,(1-z)}{m_\pi^2+2\,
    \omega^2\,(1-z)}
  +\mathcal{O}(\epsilon^4)\nonumber\\
  A_4^{\mathrm{pole}}(\omega,\,z)&=-\frac{e^2\,\omega\,(1+\kappa)^2}{2\,M^2}
  +\mathcal{O}(\epsilon^4)\nonumber\\
  A_5^{\mathrm{pole}}(\omega,\,z)&= \frac{e^2\,\omega\,(1+\kappa)^2}{2\,M^2}
  -\frac{e^2\,g_A}{8\,\pi^2\,f_\pi^2}\,\frac{\omega^3}{m_\pi^2+2\,
    \omega^2\,(1-z)}
  +\mathcal{O}(\epsilon^4)  \nonumber\\
  A_6^{\mathrm{pole}}(\omega,\,z)&=-\frac{e^2\,\omega\,(1+\kappa)}{2\,M^2}
  +\frac{e^2\,g_A}{8\,\pi^2\,f_\pi^2}\,\frac{\omega^3}{m_\pi^2+2\,
    \omega^2\,(1-z)} +\mathcal{O}(\epsilon^4)
\label{eq:pole}
\end{align}
Finally we present explicit expressions for the leading-one-loop order structure-dependent SSE Compton amplitudes including the kinematical as well as the short-distance corrections discussed in Sect.~\ref{subsec:hbchpt}.
The threshold correction was done as follows for each diagram in 
Fig.~\ref{Npicontinuum}:
If the pion propagator in a loop integral exhibits a cut at $\omega=m_\pi$, 
one replaces $\omega$ in that propagator by Eq.~(\ref{substitut}) in order to obtain the physically correct $s$-channel cut position at $\omega=\omega_\pi$. The $u$-channel contribution is unchanged. We are aware, that this procedure violates crossing symmetry, but the crossing violating effects in the $u$-channel are quite small. Formally, the terms correcting for the exact location of the pion threshold start to appear at $\mathcal{O}(p^4)$. 
%and will be corrected for in a complete next-to-leading one loop order calculation.

\newpage

\begin{equation}
\begin{split}
\bar{A}_1^H&(\omega, z)=
   \frac{b_1^2\,e^2\,\omega^2\,z}{9\,M^2}
   \left(-\frac{1}{\omega_s -\Delta_0}+\frac{1}{\omega_u +\Delta_0}\right)
  +\frac{\alpha\,(g_{118}\,t-g_{117}\,\omega^2)}{2\, \pi\,f_{\pi}^2\,M}    \\
& +\frac{\alpha}         {18\,\pi\,f_{\pi}^2   }\,
   \int\limits_0^1 dx\int\limits_0^1 dy \Biggl\{ 9\,g_A^2 \,\Biggl[ m_\pi\,\pi + 
      \frac{\pi \,\left( 2\,{m_\pi}^2 - t \right) }
           {2\,{\sqrt{-t}}}\,\arctan \left(\frac{{\sqrt{-t}}}{2\,m_\pi}\right) \\
&+ 
      \frac{\omega_s - \omega}{8\,\omega_s\,\omega}
       \,\left( {m_\pi}^2\,{\pi }^2 - 4\,\omega_s\,\omega \right) + 
      \frac{{m_\pi}^2}{2\,\omega_s\,\omega}
       \,\left( \omega\,{\arccos^2 \left(-\frac{\omega_s}{m_\pi}\right)} - 
           \omega_s\,{\arccos^2 \left(\frac{\omega}{m_\pi}\right)} \right) \\
&- 
      \left( 1 - y \right) \,\Biggl( 
         \frac{1}{c_u}\,\left[ 5\,c_u^2 - 
              \left( 1 - y \right) \,\left( \omega^2\,x^2\,\left( 1 - y \right)  
       + t\,\left( \frac{x}{2} + \left( 1 - x \right) \,y \right)  \right)  \right] \,
            \arccos \left(\frac{\omega\,x\,\left( 1 - y \right) }{{d}}\right) \\
&+ 
         \frac{1}{c_s}\,\left[ 5\,c_s^2 - \left( 1 - y \right) \,
               \left( \omega^2\,x^2\,\left( 1 - y \right)  + t\,\left( \frac{x}{2} 
     + \left( 1 - x \right) \,y \right)  \right)  \right] \,
            \arccos \left(\frac{\omega_s\,x\,\left( -1 + y \right) }{{d}}\right) 
       \Biggl)  \Biggl]\\
& +16\,g_{\pi N\Delta_0}^2\,\Biggl[ -2\,{\Delta_0}\,\ln {m_{\pi}} 
  - 3\,  {\Delta_0}\,\ln {\sqrt{m_{\pi}^2 - t\,\left( 1 - x \right) \,x}}\\
& +   {\sqrt{-m_{\pi}^2 + {\left( {\Delta_0} - \omega    \right) }^2}}\,\ln R(\Delta_0-\omega   ) 
  +   {\sqrt{-m_{\pi}^2 + {\left( {\Delta_0} + \omega    \right) }^2}}\,\ln R(\Delta_0+\omega   )\\
& -2\,{\sqrt{-m_{\pi}^2 + {\left( {\Delta_0} - \omega\,x \right) }^2}}\,\ln R(\Delta_0-\omega\,x)
  -2\,{\sqrt{-m_{\pi}^2 + {\left( {\Delta_0} + \omega\,x \right) }^2}}\,\ln R(\Delta_0+\omega\,x)\\
& -\frac{\left( 3\,{{\Delta_0}}^2 - 3\,m_{\pi}^2 + 4\,t\,\left( 1 - x \right) \,x\right)}
        {{\sqrt{   {{\Delta_0}}^2 -    m_{\pi}^2 + t\,\left( 1 - x \right) \,x}}} \,
   \ln\left(\frac{{\Delta_0}+{\sqrt{{{\Delta_0}}^2-m_{\pi}^2+t\,\left( 1 - x \right)\,x}}}
                          {{\sqrt{             m_{\pi}^2-t\,\left( 1 - x \right)\,x}}}\right)\\
& +\Biggl(\frac{1}{{C_s}}\left( 5\,C_s^2 + \omega^2\, x^2\,{\left( 1 - y \right) }^2 
  +\frac{1}{2}\,t\, x\,\left( 1 - y \right)  + t\,\left( 1 - x \right) \,\left( 1 - y \right) \,y
   \right)\\
&\times \ln \tilde{R}\left(\Delta_0-\omega\,x\,(1-y)\right) + 10\,{\Delta_0}\,\ln d \\
&+        \frac{1}{{C_u}}\left( 5\,C_u^2 + \omega^2\, x^2\,{\left( 1 - y \right) }^2 
  +\frac{1}{2}\,t\, x\,\left( 1 - y \right)  + t\,\left( 1 - x \right) \,\left( 1 - y \right) \,y
   \right)\\
& \times \ln \tilde{R}\left(\Delta_0+\omega\,x\,(1-y)\right) \Biggl)\,(1-y)  \Biggl]  \Biggl\}
  +\mathcal{O}\left(\epsilon^4\right) 
\end{split}
\label{A1}
\end{equation}

\newpage

\begin{equation}
\begin{split}
\bar{A}_2^H&(\omega, z)=
   \frac{b_1^2\,e^2\,\omega^2}{9\,M^2}
    \left(\frac{1}{\omega_s -\Delta_0}-\frac{1}{\omega_u +\Delta_0}\right)
  -\frac{\alpha\,g_{118}}{\pi\,f_{\pi}^2\,M}\,\omega^2\\
& +\frac{\alpha}         {18\,\pi\,f_{\pi}^2   }\,\int\limits_0^1 dx\int\limits_0^1 dy\,
   \omega^2\,(1-y)\,
        \Biggl\{ 9\,g_A^2\,
  \Biggl[\left( 1 - x \right)\,x \\
&\times\left( \frac{\omega_s}{c_s^2\,d^2} - 
              \frac{\omega}{c_u^2\,d^2} \right) \,{\left( 1 - y \right) }^3\,y\,
         \left( \omega^2\,x^2\,\left( 1 - y \right)  
         + t\,\left( \frac{x}{2} + \left( 1 - x \right) \,y \right)  \right)  \\
&- 
        \frac{1}{c_s^3}\biggl( \left( -1 + x \right) \,{\left( 1 - y \right) }^2\,y\,
              \left( \omega^2\,x^2\,\left( 1 - y \right)  
         + t\,\left( \frac{x}{2} + \left( 1 - x \right) \,y \right)  \right)  \\
&+ 
             c_s^2\,\left( x\,y + \left( 1 - x \right) \,
                    \left( 1 - 7\,y + 7\,y^2 \right)  \right)  \biggl) \,
           \arccos \left( \frac{\omega_s\,x\,\left( -1 + y \right) }{{d}} \right)  \\
&- 
        \frac{1}{c_u^3}\biggl( \left( -1 + x \right) \,{\left( 1 - y \right) }^2\,y\,
              \left( \omega^2\,x^2\,\left( 1 - y \right)  
         + t\,\left( \frac{x}{2} + \left( 1 - x \right) \,y \right)  \right)  \\
&+ 
             c_u^2\,\left( x\,y + \left( 1 - x \right) \,
                    \left( 1 - 7\,y + 7\,y^2 \right)  \right)  \biggl) \,
           \arccos \left(\frac{\omega\,x\,\left( 1 - y \right) }{{d}}\right) \Biggl]  \\
& -16\,g_{\pi N\Delta_0}^2\,\Biggl[ \left( 1 - x \right) \,\left( 
   \frac{-{\Delta_0} +\omega\,x\,\left( 1 - y \right) }{C_s^2\,{d}^2} -
   \frac{ {\Delta_0} +\omega\,x\,\left( 1 - y \right) }{C_u^2\,{d}^2} \right)\\
&  \times \,   {\left( 1 - y \right) }^2\,y\,\left(\omega^2\, x^2\,\left( 1-y \right)
   +\frac{1}{2}\,t\,x+t\,\left( 1 - x \right)\,y \right)  \\
& +\frac{1}{C_s^{3}}
   \biggl( C_s^2\,\left( \left( 1 - x \right) \,\left( 1 - 7\,y \right)\,
                  \left(1-y\right)+y\right)  \\
& +                               \left( 1 - x \right)\,
               {\left( 1 - y \right) }^2\,y\,\left(\omega^2\, x^2\,\left( 1-y \right)
   +\frac{1}{2}\,t\,x+t\,\left( 1 - x \right)\,y \right)  \biggl)\\
&\times\ln \tilde{R}\left(\Delta_0-\omega\,x\,(1-y)\right) 
  +\frac{1}{C_u^{3}}
   \biggl( C_u^2\,\left( \left( 1 - x \right) \,\left( 1 - 7\,y \right)\,
                  \left(1-y\right)+y\right)  \\
& + \left( 1 - x \right)\,{\left( 1 - y \right) }^2\,y\,\left(\omega^2\, x^2\,\left( 1-y \right)
  +\frac{1}{2}\,t\,x+t\,\left( 1 - x \right)\,y  \right)  \biggl)\\
&  \times\ln \tilde{R}\left(\Delta_0+\omega\,x\,(1-y)\right) \Biggl]  \Biggl\}
  +\mathcal{O}\left(\epsilon^4\right)  
\end{split}
\end{equation} 

\newpage

\begin{equation}
\begin{split}
\bar{A}_3^H&(\omega, \,z) =\frac{b_1^2\,e^2\,\omega^3\,z}{18\,M^2\,\Delta_0}
    \left(\frac{1}{\omega_s -\Delta_0}-\frac{1}{\omega_u +\Delta_0}\right)\\
& +\frac{\alpha} {\pi\,f_\pi^2}\,\int\limits_0^1 dx\int\limits_0^1 dy\,
  \Biggl\{ \frac{g_A^2}{2}\, \Biggl[-\frac{ \omega_s + \omega 
             }{8\,\omega_s\,\omega}\,\left( {m_\pi}^2\,{\pi }^2 + 4\,\omega_s\,\omega \right) \\
&+ 
      \frac{{m_\pi}^2 }{2\,\omega_s\,\omega}
      \,\left( \omega\,{\arccos^2 \left(- \frac{\omega_s}{m_\pi} \right) } + 
           \omega_s\,{\arccos^2 \left(\frac{\omega}{m_\pi}\right)} \right) \\
&+ 
      \omega^4\,\left( 1 - x \right) \,x\,{\left( 1 - y \right) }^3\,y\,\left( 1 - z^2 \right) \,
       \Biggl( \left( \frac{\omega_s}{c_s^2\,d^2} + \frac{\omega}{c_u^2\,d^2} \right) \,x\,
          \left( 1 - y \right)  \\
&
      - \frac{1}{c_u^3}
                \arccos \left(\frac{\omega\,x\,\left( 1 - y \right) }{{d}}\right)
      + \frac{1}{c_s^3}
                \arccos \left(\frac{\omega_s\,x\,\left(-1 + y \right) }{{d}}\right)
        \Biggl) \Biggl] \\
& +\frac{4\, g_{\pi N \Delta_0}^2}{9}\, \Biggl[
  -   {\sqrt{-m_\pi^2 +{\left( \Delta_0 - \omega     \right)}^2}}\,\ln R(\Delta_0-\omega   )
  +   {\sqrt{-m_\pi^2 +{\left( \Delta_0 + \omega     \right)}^2}}\,\ln R(\Delta_0+\omega   )\\
& +2\,{\sqrt{-m_\pi^2 +{\left( \Delta_0 - \omega\, x \right)}^2}}\,\ln R(\Delta_0-\omega\,x)
  -2\,{\sqrt{-m_\pi^2 +{\left( \Delta_0 + \omega\, x \right)}^2}}\,\ln R(\Delta_0+\omega\,x)\\
& -\omega^4\, \left(1 - x \right)\, x\, {\left(1 - y \right)}^3\, y\, \left(1 - z^2 \right)\,
   \Biggl( 
   \frac{\Delta_0 - \omega\, x\, \left(1 - y \right)}{C_s^2\, d^2} 
  -\frac{\Delta_0 + \omega\, x\, \left(1 - y \right)}{C_u^2\, d^2} \\
& -\frac{1}{C_s^{3}} \ln \tilde{R}\left(\Delta_0-\omega\,x\,(1-y)\right)
  +\frac{1}{C_u^{3}} \ln \tilde{R}\left(\Delta_0+\omega\,x\,(1-y)\right) \Biggl) \Biggl]\Biggl\}
  +\mathcal{O}(\epsilon^4)
\end{split}
\end{equation}

\begin{equation}
\begin{split}
\bar{A}_4^H&(\omega, \, z) =\frac{b_1^2\,e^2\,\omega^3}{18\,M^2\,\Delta_0}
    \left(\frac{1}{\omega_s -\Delta_0}-\frac{1}{\omega_u +\Delta_0}\right) \\
& +\frac{\alpha}{\pi\,f_\pi^2} \,\int\limits_0^1 dx\int\limits_0^1 dy\,\omega^2\,x\,(1-y)^2\\
&\times
   \Biggl\{ \frac{g_A^2}{2}\,
   \left[ - \frac{1}{{c_u}}
              \arccos \left(\frac{\omega\,x\,\left( 1 - y \right) }{{d}}\right)+ 
              \frac{1}{{c_s}}
              \arccos \left(\frac{\omega_s\,x\,\left(-1 + y \right) }{{d}}\right)\right] \\
& +\frac{4\, g_{\pi N \Delta_0}^2}{9}\, \left[
  -\frac{1}{{C_s}}\ln \tilde{R}\left(\Delta_0-\omega\,x\,(1-y)\right) 
  +\frac{1}{{C_u}}\ln \tilde{R}\left(\Delta_0+\omega\,x\,(1-y)\right) \right] \Biggl\}
  +\mathcal{O}(\epsilon^4)
\end{split}
\end{equation}

\newpage

\begin{equation}
\begin{split}
\bar{A}_5^H&(\omega, \, z) =\frac{b_1^2\,e^2\,\omega^3}{18\,M^2\,\Delta_0}
    \left(-\frac{1}{\omega_s -\Delta_0}+\frac{1}{\omega_u +\Delta_0}\right)\\ 
& +\frac{\alpha}{\pi\,f_\pi^2} \,\int\limits_0^1 dx\int\limits_0^1 dy\,
   \omega^2\,(1-y)\,y\,
   \Biggl\{ \frac{g_A^2}{2}\,\Biggl[ \omega^2\,\left( \frac{\omega_s}{c_s^2\,d^2} 
                                                    + \frac{\omega  }{c_u^2\,d^2} \right) \,
       \left( 1 - x \right) \,x^2\,{\left( 1 - y \right) }^3\,z \\
&- 
      \frac{1}{c_u^3}
      \left( -c_u^2 + \omega^2\,\left( 1 - x \right) \,x\,{\left( 1 - y \right) }^2\,z \right) \,
      \arccos \left(\frac{\omega\,x\,\left( 1 - y \right) }{{d}}\right) \\
&+ 
      \frac{1}{c_s^3}
      \left( -c_s^2 + \omega^2\,\left( 1 - x \right) \,x\,{\left( 1 - y \right) }^2\,z \right) \,
      \arccos \left(\frac{\omega_s\,x\,\left(-1 + y \right) }{{d}}\right)\Biggl]\\
& +\frac{4\, g_{\pi N \Delta_0}^2}{9}\, \Biggl[ 
   \frac{1}{{C_s}}\ln \tilde{R}\left(\Delta_0-\omega\,x\,(1-y)\right) 
  -\frac{1}{{C_u}}\ln \tilde{R}\left(\Delta_0+\omega\,x\,(1-y)\right) \\
& -\omega^2\, \left(1 - x \right)\, x\, {\left(1 - y \right)}^2\, z\,\Biggl( 
   \frac{\Delta_0 - \omega\, x\, \left(1 - y \right)}{C_s^2\, d^2} 
  -\frac{\Delta_0 + \omega\, x\, \left(1 - y \right)}{C_u^2\, d^2} \\
& -\frac{1}{C_s^{3}}\ln \tilde{R}\left(\Delta_0-\omega\,x\,(1-y)\right) 
  +\frac{1}{C_u^{3}}\ln \tilde{R}\left(\Delta_0+\omega\,x\,(1-y)\right) \Biggl) \Biggl] \Biggl\}
  +\mathcal{O}(\epsilon^4)
\end{split}
\end{equation}

\begin{equation}
\begin{split}
\bar{A}_6^H&(\omega, \, z) = 
   \frac{\alpha}{\pi\,f_\pi^2} \,\int\limits_0^1 dx\int\limits_0^1 dy\,\omega^2\,(1-y)\,y\,  
   \Biggl\{ \frac{g_A^2}{2}\,\Biggl[- \omega^2\,\left( \frac{\omega_s }{c_s^2\,d^2} + 
         \frac{\omega }{c_u^2\,d^2} \right) \,\left( 1 - x \right) \,x^2\,
        {\left( 1 - y \right) }^3 \\
&+ 
      \frac{1}{c_u^3}
      \left( -c_u^2 + \omega^2\,\left( 1 - x \right) \,x\,{\left( 1 - y \right) }^2 \right) \,
      \arccos \left(\frac{\omega\,x\,\left( 1 - y \right) }{{d}}\right) \\
&- 
      \frac{1}{{c_s^3}}
      \left( -c_s^2 + \omega^2\,\left( 1 - x \right) \,x\,{\left( 1 - y \right) }^2 \right) \,
      \arccos \left(\frac{\omega_s \,x\,\left( -1 + y \right) }{{d}}\right)\Biggl]\\
& +\frac{4\, g_{\pi N \Delta_0}^2}{9}\, \Biggl[
  -\frac{1}{{C_s}}\ln \tilde{R}\left(\Delta_0-\omega\,x\,(1-y)\right)
  +\frac{1}{{C_u}}\ln \tilde{R}\left(\Delta_0+\omega\,x\,(1-y)\right) \\ 
& +\omega^2\, \left(1 - x \right)\, x\, {\left(1 - y \right)}^2\,    \Biggl( 
   \frac{\Delta_0 - \omega\, x\, \left(1 - y \right)}{C_s^2\, d^2} 
  -\frac{\Delta_0 + \omega\, x\, \left(1 - y \right)}{C_u^2\, d^2} \\
& -\frac{1}{C_s^{3}}\ln \tilde{R}\left(\Delta_0-\omega\,x\,(1-y)\right) 
  +\frac{1}{C_u^{3}}\ln \tilde{R}\left(\Delta_0+\omega\,x\,(1-y)\right) \Biggl) \Biggl] \Biggl\}
  +\mathcal{O}(\epsilon^4)
\end{split}
\label{A6}
\end{equation}

\newpage

In Eqs.~(\ref{A1})--(\ref{A6}) we have used the following abbreviations:

\begin{align*}
       d   ^2  &=m_\pi^2 - t\,\left( 1 - x \right) \,\left( 1 - y \right) \,y\\
       c_s^2   &=d^2-\omega_s^2\,x^2\,(1-y)^2\\
       c_u^2   &=d^2-\omega  ^2\,x^2\,(1-y)^2\\
       C_s ^2  &=({\Delta_0}   - \omega     \,x  \,    \left( 1 - y \right))^2 - d^2     \\
       C_u ^2  &=({\Delta_0}   + \omega     \,x  \,    \left( 1 - y \right))^2 - d^2     \\
\\
       \omega_s&=\sqrt{s}-M\\
       \omega_u&=M-\sqrt{u}\\
       s       &=\left(p+k \right)^2=\left(\omega+\sqrt{M^2+\omega^2}\right)^2\\
       t       &=\left(k-k'\right)^2=2\,\omega^2\,(z-1)\\
       u       &=\left(p-k'\right)^2=M^2-2\,\omega\,\sqrt{M^2+\omega^2}-2\,\omega^2\,z\\
\\
       R (X)   &=\frac{X}{m_\pi}+\sqrt{\frac{X^2}{m_\pi^2}-1},\;\,
\tilde{R}(X)    =\frac{X}{d    }+\sqrt{\frac{X^2}{d    ^2}-1}  \\
\end{align*}

For the \emph{isovector} Compton structure amplitudes one
finds a null result to leading-one-loop order:
\begin{eqnarray}
\bar{A}_i^{H\,(v)}&=&0+{\cal O}(\epsilon^4)\;,
\end{eqnarray}
with $i=1,\dots, 6$.

\newpage

%%%%%%%%%%%%%%% Intro %%%%%%%%%%%%%%%%%%%
\section{Projection Formulae for $\chi$EFT}
\setcounter{equation}{0}
\label{sec:AppC}
%%%%%%%%%%%%%%%%%%%%%%%%%%%%

The connection between the Compton structure amplitudes
$\bar{A}_i^H(\omega,z),\;i=1,\ldots, 6$ given in the previous section and the cm Compton multipoles $f_{XX'}^{l\pm}(\omega),\;X,X'=E,M$, introduced in Sect.~\ref{sec:multipoles}, reads:
\begin{align*}
\begin{split}
  f_{EE}^{1+}(\omega)&=\int\limits_{-1}^{1}\frac{M}{16\cdot 4\pi\,W}
  \biggl[\bar{A}_3^H(\omega,z)\left(-3+z^2\right)+4 \bar{A}_6^H(\omega,z)
  \left(-1+z^2\right)\\
  &+\left(2
    \bar{A}_2^H(\omega,z)+\bar{A}_4^H(\omega,z)+2\,\bar{A}_5^H(\omega,z)\right)
  \,z\,\left(-1+z^2\right)+2 \bar{A}_1^H(\omega,z)\left(1+z^2\right)\biggl]dz\\
\end{split}\\
\begin{split}
  f_{EE}^{1-}(\omega)&=\int\limits_{-1}^{1}\frac{M}{8\cdot 4\pi\,W}
  \biggl[-\bar{A}_3^H(\omega,z)\left(-3+z^2\right)-4 \bar{A}_6^H(\omega,z)
  \left(-1+z^2\right)\\
  &-\left(-\bar{A}_2^H(\omega,z)+\bar{A}_4^H(\omega,z)+2\,\bar{A}_5^H(\omega,z)
  \right)
  \,z\,\left(-1+z^2\right)+\bar{A}_1^H(\omega,z)\left(1+z^2\right)\biggl]dz\\
\end{split}\\
\begin{split}
  f_{MM}^{1+}(\omega)&=\int\limits_{-1}^{1}\frac{M}{16\cdot 4\pi\,W}
  \biggl[2 \bar{A}_2^H(\omega,z)\left(-1+z^2\right)\\
  &+\bar{A}_4^H(\omega,z)\left(-1+z^2\right)+2\left(\bar{A}_5^H(\omega,z)
    \left(1-z^2\right)
    +\bar{A}_1^H(\omega,z)\,2 z-\bar{A}_3^H(\omega,z)\,z\right)\biggl]dz\\
\end{split}\\
\begin{split}
  f_{MM}^{1-}(\omega)&=\int\limits_{-1}^{1}\frac{M}{8\cdot 4\pi\,W}
  \biggl[\bar{A}_4^H(\omega,z)\left(1-z^2\right)\\
  &+\bar{A}_2^H(\omega,z)\left(-1+z^2\right)+2\left(\bar{A}_5^H(\omega,z)
    \left(-1+z^2\right)+\bar{A}_1^H(\omega,z)\,z
    +\bar{A}_3^H(\omega,z)\,z\right)\biggl]dz\\
\end{split}\\
\begin{split}
  f_{EE}^{2+}(\omega)&=\int\limits_{-1}^{1}\frac{M}{72\cdot 4\pi\,W}
  \biggl[\bar{A}_4^H(\omega,z)\left(-1-3z^2+4z^4\right)
  +\bar{A}_2^H(\omega,z)\left(3-9z^2+6z^4\right)+2(\bar{A}_5^H(\omega,z)\\
  &\times\left(-1-3z^2+4z^4\right)+\bar{A}_1^H(\omega,z)\,3z^3
    +\bar{A}_3^H(\omega,z)\left(2z^3-3z\right)+\bar{A}_6^H(\omega,z)\left(6z^3-6z
    \right))\biggl]dz\\
\end{split}\\
\begin{split}
  f_{EE}^{2-}(\omega)&=\int\limits_{-1}^{1}\frac{M}{48\cdot 4\pi\,W}
  \biggl[\bar{A}_4^H(\omega,z)\left(1+3z^2-4z^4\right)+\bar{A}_2^H(\omega,z)
  \left(2-6z^2+4z^4\right)+2(\bar{A}_5^H(\omega,z)\\
  &\times\left(1+3z^2-4z^4\right)+\bar{A}_1^H(\omega,z)\,2z^3
    +\bar{A}_3^H(\omega,z)\left(3z-2z^3\right)+\bar{A}_6^H(\omega,z)\left(6z-6z^3
    \right))\biggl]dz\\
\end{split}
\end{align*}
\newpage
\begin{align}
\begin{split}
  f_{MM}^{2+}(\omega)&=\int\limits_{-1}^{1}\frac{M}{72\cdot 4\pi\,W}
  \biggl[\bar{A}_3^H(\omega,z)\left(1-3z^2\right)\\
  &+\left(3 \bar{A}_2^H(\omega,z)+5 \bar{A}_4^H(\omega,z)-2
    \bar{A}_5^H(\omega,z)\right)
  \,z\,\left(-1+z^2\right)+\bar{A}_1^H(\omega,z)\left(-3+9 z^2\right)\biggl]dz\\
\end{split}\nonumber\\
\begin{split}
  f_{MM}^{2-}(\omega)&=\int\limits_{-1}^{1}\frac{M}{48\cdot 4\pi\,W}
  \biggl[\bar{A}_3^H(\omega,z)\left(-1+3 z^2\right)\\
  &+\left(2 \bar{A}_2^H(\omega,z)-5 \bar{A}_4^H(\omega,z)+2
    \bar{A}_5^H(\omega,z)\right)
  \,z\,\left(-1+z^2\right)+\bar{A}_1^H(\omega,z)\left(-2+6 z^2\right)\biggl]dz\\
\end{split}\nonumber\\
\begin{split}
  f_{EM}^{1+}(\omega)&=\int\limits_{-1}^{1}\frac{M}{16\cdot 4\pi\,W}
  \biggl[\bar{A}_3^H(\omega,z)\left(1-3z^2\right)\\
  &-2 \bar{A}_6^H(\omega,z)\left(-1+z^2\right)-\left(\bar{A}_4^H(\omega,z)+4
    \bar{A}_5^H(\omega,z)\right)
  \,z\,\left(-1+z^2\right)\biggl]dz\\
\end{split}\nonumber\\
\begin{split}
  f_{ME}^{1+}(\omega)&=\int\limits_{-1}^{1}\frac{M}{16\cdot
    4\pi\,W}\biggl[\bar{A}_4^H(\omega,z)\left(1-z^2\right)
  -2z\,\left(\bar{A}_3^H(\omega,z)+\bar{A}_6^H(\omega,z)\left(1-z^2\right)\right)\biggl]dz\\
\end{split}
\end{align} 

%%%%%%%%%%%%%%%%%%%%%%%%%%%%%%%%%%%%%%
\section{Static Quadrupole-Polarizabilities}
\label{sec:quadrupole}
%%%%%%%%%%%%%%%%%%%%%%%%%%%%%%%%%%%%%

The spin-independent static quadrupole polarizabilities have been analyzed in Ref.~\cite{holstein} to leading-one-loop order in SSE. Here we present the details of our results for these $l=2$ polarizabilities, as they turn out to be quite different: 
\begin{align}
\begin{split}
  \bar{\alpha}_{E2}&= \frac{ \alpha\,g_A^2}{ 32\,F_\pi^2\, m_\pi^3\,\pi }
  \left(\frac{7}{5}+\frac{ 9}{10}\,\frac{m_\pi}{M}\,\frac{1}{\pi}\right)
  \nonumber\\
  & +\frac{ \alpha\,g_{\pi N \Delta_0}^2}{135\,(F_\pi\,\pi)^2\, m_\pi^2 } 
   \left[\frac{\Delta_0\,\left(11\,\Delta_0^2-41\,m_\pi^2\right)}
     {\left(\Delta_0^2-m_\pi^2\right)^2}
    +\frac{3\,m_\pi^2\,\left(3 \,\Delta_0^2+
        7\,m_\pi^2\right)}{\left(\Delta_0^2-m_\pi^2\right)^{5/2}}\ln R\right]+{\cal O}(\epsilon^4)
\end{split}\nonumber\\
&= \left[ 21.48\,(N\pi)+0\,(c.t.)+0\,(\Delta-pole)+4.99\,(\Delta\pi)\right]
\times 10^{-4}\;\mathrm{fm}^5\nonumber\\
&=\left[26.47+{\cal O}(\epsilon^4)\right]\times 10^{-4}\;\mathrm{fm}^5\nonumber\\
\begin{split}
  \bar{\beta}_{M2} &=-\frac{ 3\,\alpha\,g_A^2}{ 160\,F_\pi^2\, m_\pi^3\,\pi }
  -\frac{2\,\alpha\,b_1^2}{3\,\Delta_0^2\,M^3}  
\nonumber\\
  & +\frac{ \alpha\,g_{\pi N \Delta_0}^2}{ 15\,(F_\pi\,\pi)^2 \,m_\pi^2 }
  \left[\frac{-\Delta_0}{\Delta_0^2-m_\pi^2}+\frac{m_\pi^2}{(\Delta_0^2-m_\pi^2
      )^{3/2}}\ln R\right]+{\cal O}(\epsilon^4)
\end{split}\nonumber\\
&= \left[-8.93\,(N\pi)+ 0\,(c.t.)-(5.18\pm0.32)
   \,(\Delta-pole)-3.37\,(\Delta\pi)\right]\times 10^{-4}\;\mathrm{fm}^5
   \nonumber\\
&=\left[-17.48\pm0.32+{\cal O}(\epsilon^4)\right]\times 10^{-4}\;\mathrm{fm}^5
\end{align}
$ R=\left(\Delta_0+\sqrt{\Delta_0^2-m_\pi^2}\right)/{m_\pi}$ is a dimension-less parameter \cite{HHK97}. In particular we note the extra piece $\sim m_\pi^{-2}$ from the one-pion threshold correction in $\bar{\alpha}_{E2}$ as well as the kinematically induced $u$-channel $\Delta$(1232) contribution in $\bar{\beta}_{M2}$. For details on the origin of these terms we refer to Sect.~\ref{subsec:hbchpt}, items 1 and 2. Judging from the plots of the corresponding dynamical quadrupole polarizabilities as shown in Fig.~\ref{fig3}, each of these effects seems to improve the agreement between SSE and Dispersion Theory. However, as discussed in Sect.~\ref{subsec:experiment}, we remind the reader that the $l=2$ polarizabilities are in effect so small, that they cannot be determined from state-of-the-art nucleon Compton scattering experiments.

%%%%%%%%%%%%%%%%%%%%%%%%%%%%%%%%%%%%%%%%%%%%%%%%%%%%%%%%%%%%%%%%%%%%%%%%%%%%%%%
%%%%%%%%%%%%%%%%%%%%%%%%%%%%%%%%%%%%%%%%%%%%%%%%%%%%%%%%%%%%%%%%%%%%%%%%%%%%%%%
\newpage
%%%%%%%%%%%%%%%%%%%%%%%%%%%%%%%%%%%%%%%%%%%%%%%%%%%%%%%%%%%%%%%%%%%%%%%%%%%%%%%

%%%%%%%%%%%%%%%%%%%%%%%%%%%%%%%%%%%%%%%%%%%%%%%%%%%%%%%%%%%%%%%%%%%%%%%%%%

%%%%%%%%%%%%%%%%%%%%%%%%%%%%%%%%%%%%%%%%%%%%%%%%%%%%%%%%%%%%%%%%%%%%%%%%%%

\end{document}